\documentclass[a4paper,fleqn,usenatbib]{mnras}

\usepackage{newtxtext,newtxmath}

\usepackage[T1]{fontenc}
\usepackage{ae,aecompl}

\usepackage{graphicx}	
\usepackage{amsmath}	

\title[Porosity of interstellar dust]{Evolution of dust porosity through
coagulation and shattering in the interstellar medium}

\author[H. Hirashita]{
Hiroyuki Hirashita,$^1$\thanks{E-mail: hirashita@asiaa.sinica.edu.tw}
Vladimir B. Il'in,$^{2,3,4}$
Laurent Pagani$^5$ and Charl\`{e}ne Lef\`{e}vre$^{6,5}$
\\
$^{1}$Institute of Astronomy and Astrophysics, Academia Sinica,
Astronomy-Mathematics Building, No.\ 1, Section 4,\\
Roosevelt Road, Taipei 10617, Taiwan\\
$^{2}$St Petersburg State University, Universitetskij Pr. 28, St Petersburg 198504, Russia\\
$^3$Pulkovo Observatory, Pulkovskoe Sh. 65/1, St Petersburg 196140, Russia\\
$^4$St Petersburg University of Aerospace Instrumentation, Bol. Morskaya 67, St Petersburg 190000, Russia\\
$^5$LERMA \& UMR8112 du CNRS, Observatoire de Paris, PSL University,
Sorbonne Universit\'{e}s,  CNRS, F-75014 Paris, France\\
$^6$Institut de Radioastronomie Millim\'{e}trique (IRAM), 300 rue de la Piscine, 38400 Saint-Martin
d'H\`{e}res, France
}

\date{Accepted XXX. Received YYY; in original form ZZZ}

\pubyear{2020}

\begin{document}
\label{firstpage}
\pagerange{\pageref{firstpage}--\pageref{lastpage}}
\maketitle

\begin{abstract}
The properties of interstellar grains, such as grain size distribution and
grain porosity, are affected by interstellar processing, in particular,
coagulation and shattering, which take place in the dense and diffuse interstellar medium (ISM),
respectively.
In this paper, we formulate and calculate the evolution of grain size distribution and
grain porosity through shattering and coagulation.
For coagulation, we treat the grain evolution depending on the collision energy.
Shattering is treated as
a mechanism of forming small compact fragments.
The balance between these processes are determined by the dense-gas mass fraction
$\eta_\mathrm{dense}$, which determines the time fraction of coagulation relative to
shattering. We find that
the interplay between shattering
supplying small grains and coagulation forming porous grains from shattered grains
is fundamentally important in creating and maintaining porosity. 
The porosity rises to 0.7--0.9 (or the filling factor 0.3--0.1) around grain radii
$a\sim 0.1~\micron$.
We also find that, in the case of $\eta_\mathrm{dense}=0.1$ (very efficient shattering with
weak coagulation) porosity significantly enhances coagulation, creating fluffy submicron
grains with filling factors lower than 0.1.
The porosity enhances the extinction by 10--20 per cent at all wavelengths for amorphous carbon and
at ultraviolet wavelengths for silicate. The extinction curve shape of
silicate becomes steeper if we take porosity into account.
We conclude that the interplay between shattering and
coagulation is essential in creating porous grains in the interstellar medium and that
the resulting porosity can impact the grain size distributions and extinction curves.
\end{abstract}


\section{Introduction}\label{sec:intro}

Dust grains are important for various processes in the
interstellar medium (ISM) and in galaxies. Not only the total dust abundance but
also the distribution function of grain radius, referred to as the grain size distribution,
is important, since under a given dust abundance, the total surface area and the
cross-section of light absorption and scattering (extinction) are governed by
the grain size distribution. Since H$_2$ forms on dust surfaces
\citep[e.g.][]{Gould:1963aa,Cazaux:2004aa}, the total surface area of dust grains affects the H$_2$
formation rate, which could influence the galaxy evolution
\cite[e.g.][]{Yamasawa:2011aa,Chen:2018aa}.
The extinction cross-section as a function of wavelength, referred to as the
extinction curve, gives us a clue to the grain size distribution
\citep[e.g.][hereafter MRN]{Mathis:1977aa}. The extinction curve is strongly modified by dust evolution
\cite[e.g.][]{Asano:2014aa}.
The evolution of grain size distribution also affects the
spectral energy distributions (SEDs) of galaxies \citep[e.g.][]{Desert:1990aa,Takeuchi:2005aa}.
Thus, the evolution of grain size distribution is of fundamental importance in understanding
the chemical and radiative processes in the ISM and in galaxies.

The evolution of dust in the ISM or in galaxies is governed by various processes
\citep[e.g.][]{Lisenfeld:1998aa,Draine:2009ab}.
The major processes for the dust enrichment are stellar dust production and
dust growth in the ISM \citep[e.g.][]{Dwek:1998aa}.
Dust destruction in supernova shocks \citep[e.g.][]{McKee:1989aa}
is considered to dominate the decrease of dust in the ISM. In the Milky Way environment,
since the metallicity is high enough, dust growth by the accretion of gas-phase metals
can be efficient enough to balance the dust destruction
\citep[e.g.][]{Draine:1990aa,Dwek:1998aa,Hirashita:1999aa,Inoue:2011aa,Mattsson:2014aa}.
This accretion process, {which is also verified in some experiments
\citep{Rouille:2014aa,Rouille:2020aa}}, occurs in the dense ISM.
The grain size distribution, on the other hand, is strongly modified by coagulation in the
dense ISM and
shattering in the diffuse ISM \citep[e.g.][]{ODonnell:1997aa,Hirashita:2009ab}.
Although some of the above processes occur either in a certain phase of
the ISM, all dust processes eventually affect the mean properties of interstellar
dust through a quick ($\sim 10^7$ yr) exchange between various ISM phases
\citep[e.g.][]{McKee:1989aa}.
The abundance ratio between
large and small grains is roughly determined by the balance between coagulation and
shattering \citep{Hirashita:2019aa}.

If coagulation and shattering govern the functional shape of grain size distribution as
suggested by previous studies, these processes could put a clear imprint on the
grain properties. In particular, coagulation could form inhomogeneous grains
in terms of the composition and the shape.
Such a possibility is modelled by \citet{Mathis:1989aa}, who constructed a
model of interstellar grains that are
assumed to be aggregates of multiple species containing vacuum.
They succeeded in obtaining a model that is consistent with the observed Milky Way extinction
curves. Although composite aggregates (composed of multiple grain species) could be
disrupted into single species in the diffuse ISM \citep{Li:1997aa,Hoang:2019aa},
it is at least expected that coagulation develops non-sphericity with vacuum,
which could lead to porous (or fluffy) grains.

Indeed,
fluffy large grains are observationally indicated by the X-ray scattering halos surrounding point sources,
such as X-ray binaries \citep{Woo:1994aa} and Nova Cygni 1992 \citep{Mathis:1995aa}.
\citet{Mathis:1995aa} attempted to reproduce the X-ray halo strength of Nova Cygni 1992
under the observed optical extinction towards that object.
\citet{Smith:1998aa} reconsidered the treatment of grain optical properties using the Mie solution,
and concluded, contrary to the above conclusion, that compact silicate and carbon grains
are consistent with the observation of the X-ray scattering halo around Nova Cygni 1992.
\citet{Draine:2003aa} also indicated that a compact-grain model based on \citet{Weingartner:2001aa}
is consistent with
the dust towards Nova Cygni 1992 if a lower optical extinction is adopted.
\citet{Draine:2003ab} also discussed the X-ray halo of Cyg X-1 in
addition to that of Nova Cygni 1992 and supported the same compact-grain model; however,
they also mentioned the uncertainty arising from
the distribution of dust  in the line of sight. Within the uncertainty, grains with moderate porosities
could still exist in the diffuse ISM

Considering porous grains {was} also motivated by the severe constraint on the available metals
(especially carbon) {historically}.
\citet{Mathis:1996aa} investigated a possibility of reproducing the Milky Way extinction curve
with fluffy grains. As a consequence, they `saved'
the heavy elements the most (i.e.\ used the minimum amount of metals for dust)
by including 25--65 per cent of vacuum in the dust grains.
Very fluffy grains are excluded because their extinction (absorption + scattering)
per unit dust mass is too low. However,
\citet{Dwek:1997aa} argued that, if we take the fitting to the infrared dust emission SED into account,
\citet{Mathis:1996aa}'s model overproduces the optical--ultraviolet (UV) extinction.
{\citet{Li:2005aa} showed, based on the argument on the Kramers-Kronig relations,
that the observed optical--UV extinction is still underproduced with the Galactic metal abundance
derived from B stars. We should also keep in mind that recent dust models with updated
extinction-to-column density ratios
($A_V/N_\mathrm{H}$) and the protosolar abundance
with a correction for
Galactic chemical evolution after the Sun formation indicate that
the above severe metallicity constraint may not be an issue any more
\citep{Draine:2020aa,Hensley:2020aa,Zuo:2020aa}.}

{Clarifying how porous the interstellar dust could be
is in itself a problem independent of the metal abundance constraint.}
Since the above mentioned studies did not necessarily give a definite constraint on the grain porosity,
it is still meaningful to investigate the effect of porosity on the extinction
curves in a general context.
\citet{Voshchinnikov:2005aa,Voshchinnikov:2006aa} studied the
dependence of extinction curve on the porosity. These studies give a basis on which
we constrain the allowed ranges of porosity through comparison with observations.
In reality, porosity could depend on the grain size, because the evolution of
grain size distribution plays an important role in determining the
grain porosity.
However, it is difficult to relate grain size and porosity, since
the physical link between these two quantities is still missing.
Because of this `missing link', the porosity still remains to be a free parameter.
It is definitely necessary to understand the interplay between grain size and porosity in
the evolution of interstellar dust, if we would like to obtain a meaningful insight into the
physical mechanisms of dust evolution through the observationally constrained grain porosities.

The evolution of porous grains by coagulation has been modelled and discussed in
the field of protoplanetary discs.
\citet{Ormel:2007aa} developed a Monte Carlo method for the evolution of grain size distribution
by coagulation in protoplanetary
discs taking the porosity evolution into account. In each collision, depending on the
collision energy, they treated the difference between the hit-and-stick (sticking without
modifying the grain shapes) and compaction
(decrease of grain volume by compression) 
regimes.
\citet[][hereafter O09]{Okuzumi:2009aa} formulated this problem by 
using the 2-dimensional (2D) Smoluchowski equation that treats
distribution function of two variables, grain radius and porosity. Since directly solving the 2D
equation is computationally expensive, they only concentrated on the mean porosity
at each grain radius while they fully solved the grain size distribution
\citep[see also][hereafter O12]{Okuzumi:2012aa}.
They finally obtained a set of moment equations for the grain volume: the zeroth order
describes the grain size distribution and the first order the `grain volume distribution'.
A set of grain radius and grain volume gives information on the grain porosity.
The calculated porosity is also useful to predict the radiative properties of dust grains
in protoplenetary discs \citep[e.g.][]{Kataoka:2014aa,Tazaki:2016aa}.

The above methods to trace the evolution of porous grains
are applied to the grain evolution in dense molecular clouds.
\citet{Ormel:2009aa} applied the above mentioned Monte Carlo method (but including
also grain fragmentation) and showed that the filling factor of grain could
become as small as $\sim 0.1$ (i.e.\ the fraction of vacuum is $\sim 0.9$).
Since the optical properties of dust could be affected by grain porosity,
modelling of porosity is important in interpreting the wavelength dependence of
absorption and emission in dense molecular clouds
\citep{Ormel:2011aa,Lefevre:2020aa}.
However, these works mainly consider dense ($\gtrsim 10^4$ cm$^{-3}$)
environments, and the relation to the grains actually contributing to the
interstellar extinction is not clear. As mentioned above, shattering is also
important as a modifying mechanism of grain size distribution. The functional shape
of the grain size distribution could be determined by the balance between shattering
and coagulation. This balance could also be important to determine the
porosities at various grain sizes.

The goal of this paper is to clarify the evolution of porous grains in the ISM
by considering both coagulation and shattering.
We have already formulated the evolution of grain size distribution using the Smoluchowski
equation in our previous works \citep[e.g.][]{Hirashita:2009ab,Asano:2013aa,Hirashita:2019aa}.
For the porosity evolution, we utilize the framework developed by O09 and O12
for protoplanetary discs with some modifications mainly to include shattering.
Our model developed in this paper is aimed
at being used to calculate the dust evolution on galaxy scales in the future.
Although the Smoluchowski equation usually refers to the one describing coagulation,
we broadly use this name also for an extended version for shattering.

As a result of the modelling in this paper, we will be able to predict the expected
grain porosity as a function
of grain radius in the ISM. We emphasize that the previous models of grain porosity
introduced above mostly treated porosity as a free parameter to be fitted to
observed dust properties such as extinction curves.
Thus, this study will provide the first prediction on the
grain-radius-dependent porosity expected theoretically from coagulation and shattering.
We also calculate extinction curves to clarify the effect of predicted porosity on an
observable dust property. This paper is not aimed at complete modelling but focused on
the construction of a basic framework for clarifying how large the porosity could be
in the presence of coagulation and shattering.

This paper is organized as follows.
In Section~\ref{sec:model}, we formulate the evolution of grain size distribution and porosity.
We show the results in Section~\ref{sec:result}, and present the effects on extinction curves
in Section~\ref{sec:ext_result}.
We provide some extended discussions, especially focusing on the effects of porosity
in Section \ref{sec:discussion}.
We give the conclusion of this paper in Section \ref{sec:conclusion}.

\section{Model}\label{sec:model}

First of all, we clarify some terms. The \textit{filling factor} of a grain is defined as the
volume fraction occupied by the material composing the grain. The rest
(unity minus the filling factor) is referred to as the \textit{porosity}, which is the volume
fraction of vacuum. A grain is \textit{compact} if the filling factor is unity
(i.e.\ no porosity).
In our previous papers \citep[e.g.][]{Hirashita:2019aa}
and others \citep[e.g.][]{Jones:1994aa,Jones:1996aa,Mattsson:2016aa}, the evolution of grain size
distribution in the ISM is formulated for compact grains.
Here, we extend the formulation to
include the evolution of grain porosity.

{As mentioned in the Introduction, we}
concentrate on the collisional processes for grain evolution, namely
coagulation and shattering, since they are the main
drivers for the evolution of both grain size distribution and grain porosity.
{We neglect dust evolution processes other than coagulation and shattering. This
treatment assumes that coagulation and/or shattering occur in a `closed box' without any
supply from stellar dust production and from accretion of gas-phase metals, and without
any loss of dust by supernova shock destruction.}
For simplicity, we consider a single dust species in solving the evolution of grain
size distribution and porosity to avoid the complexity arising from compound species.
That is, we concentrate on the effect of porosity in a single dust species.

Coagulation and shattering occur in the different ISM phases (dense and diffuse
ISM phases, respectively).
However, since the exchange of these phases occurs on a short time-scale
($\sim 10^7$~yr; \citealt{McKee:1989aa}), we consider the mixture of coagulation and shattering
in addition to treating each process separately. For the dense and diffuse ISM phases,
we consider representative phases that could affect the major dust populations in the ISM
and the physical conditions are given in Section \ref{subsec:param}.
The grain motions are assumed to be driven by interstellar turbulence.
We do not trace coagulation in a star-forming cloud with densities $\gtrsim 10^4$ cm$^{-3}$,
since, as mentioned in the Introduction, the relation of the grains formed in such extreme
environments to the general ISM grains is not clear. Moreover, we need to consider
physical processes such as ambipolar diffusion in considering the grain motion in star-forming
clouds \citep[e.g.][]{Silsbee:2020aa,Guillet:2020aa}.
Therefore, we leave the detailed studies of dense star-forming clouds for a future work.

{Coagulation in the dense ISM could be complex since accretion of gas-phase
materials could occur at the same time \citep{Hirashita:2014ab,Voshchinnikov:2014aa}.
Accretion could decrease the porosity because part of
the vacuum could be filled \citep[][appendix B]{Li:2003aa}.
As a result of accretion in an environment where UV radiation is shielded, ice mantles could
also develop on the grain surfaces. However, such mantles are evaporated in the diffuse ISM
so that aggregates attached through the ice mantles could dissolve once they are
injected into the diffuse ISM.
As mentioned later, whether or not ice mantles enhance coagulation is still debated, and
refractory surfaces could be more sticky (Section \ref{subsubsec:coag}).
Therefore, it is hard to clarify with the currently available knowledge how
ice mantles affect the observed grain properties (e.g.\ extinction curves) in the ISM
through coagulation.
Since we are interested in the general population of grains in the ISM
(especially, silicate and carbonaceous dust), we neglect the formation of volatile mantles,
and assume that the collisions between refractory grains in the dense ISM lead to
formation of refractory
coagulated grains. We also neglect the possible development of core-mantle structures as a
result of accretion, which could
explain the Milky Way dust properties \citep{Li:1997aa,Jones:2017aa}.
We simply examine the effect of porosity on the extinction curves of basic materials
(silicate and carbonaceous dust in this paper; Section \ref{subsec:ext_method}),
but do not aim at reproducing the observed Milky Way extinction curve.}

{In this paper, we also neglect the electric charge of grains. The grain collision cross-section
could be enhanced or reduced depending on the grain charge. The charge effect is not
important for large grains because they have high kinetic energy (i.e.\ the Coulomb force does
not alter the grain motion) as commented in \citet{Hirashita:2019aa}.  The charge may be important
for coagulation of small grains, which are positively or negatively charged depending on
the physical condition of the ambient medium and radiation field \citep{Weingartner:2001ab}.
For example, in a physical condition appropriate for molecular clouds, small grains tend to
be charged negatively, while large grains positively \citep{Yan:2004aa}. This situation
could suppress the collisions between
small grains but enhance those between a small and a large grain. At the same time, the grain charge
is very sensitive to the assumed physical conditions in the dense clouds (especially, the ionization
degree and the UV radiation field intensity, which is affected by the dust attenuation).
Although it is interesting to investigate the detailed
dependence on these physical parameters,
in this paper we choose to neglect
the effect of grain charge for the simplicity of our treatment.}

\subsection{Basic framework}\label{subsec:setup}

The evolution of grain size distribution and grain porosity by coagulation is formulated by
O09 and O12. Although their main target was dust evolution in protoplanetary discs,
the generality of their  formulation allows us to apply it to interstellar dust.
We also include shattering in this paper.
The evolution of grain size distribution and grain porosity by coagulation and shattering
is described by the distribution functions of grain mass
($m$) and grain volume ($V$)\footnote{{We use the upper case
$V$ for volumes and the lower case $v$ for velocities throughout this paper.}} at time $t$, and is governed by
the 2D Smoluchowski equation. However, as noted by O09, solving the
2D Smolchowski equation requires a huge computational expense.
Therefore, we concentrate on the moment equations for $V$, but we still solve the
full distribution function for $m$.

To treat porous grains, it is convenient to introduce the
following two types of grain radius: characteristic radius ($a_\mathrm{ch}$) and mass-equivalent
radius ($a_m$). The characteristic radius is related to the volume as
$V=(4\upi /3)a_\mathrm{ch}^3$, while the mass-equivalent radius is related to the
grain mass as $m=(4\upi /3)a_m^3s$, where $s$ is the bulk material density.
For compact grains, $m=sV$, so that $a_\mathrm{ch}=a_m$. For porous grains,
$a_\mathrm{ch}>a_m$. Using these two grain radii, the filling factor, $\phi_m$, is expressed as
$\phi_m\equiv (a_m/a_\mathrm{ch})^3 (\leq 1)$
\citep[see also][]{Ormel:2009aa}, while the porosity is $1-\phi_m$.
Since it is necessary to
specify $s$, we simply adopt the same value as used in HM20 ($s=2.24$ g~cm$^{-3}$
taken from graphite),
but adopting $s=2$--3.5 g cm$^{-3}$ appropriate for interstellar grains
\citep[e.g.][]{Weingartner:2001aa} does not change our conclusion
significantly.
We vary some parameters relevant for porosity, which are more important in this paper
(Section \ref{subsec:param}). Thus, we simply use the same parameters as in HM20 for
the material properties (specifically $s$ and $Q_\mathrm{D}^\star$ introduced later)
that do not directly affect the porosity evolution.

Before presenting the basic equations, we need to introduce some quantities.
We use the grain mass distribution instead of the grain size distribution since
there is ambiguity in the grain size for porous grains.
We denote the distribution function of $m$ and $V$ at time $t$ as
$f(m,\, V,\, t)$.
The grain mass distribution at time $t$, $\tilde{n}(m,\, t)$ is defined such that
$\tilde{n}(m,\, t)\,\mathrm{d}m$ is the number density of grains whose mass is between
$m$ and $m+\mathrm{d}m$, and is related to the above distribution function as
\begin{align}
\tilde{n}(m,\, t)\equiv\int_0^\infty f(m,\, V,\, t)\,\mathrm{d}V.\label{eq:n}
\end{align}
We also introduce the first moment of $f$ for $V$ as
\begin{align}
\bar{V}(m,\, t)\equiv\frac{1}{\tilde{n}(m,\, t)}\int_0^\infty Vf(m,\, V,\, t)\,\mathrm{d}V.\label{eq:V}
\end{align}
This quantity ($\bar{V}$) is the mean volume of grains with mass $m$.
We define the following two quantities,
$\rho (m,\, t)\equiv m\tilde{n}(m,\, t)$ and $\psi (m,\, t)\equiv \bar{V}(m,\, t)\tilde{n}(m,\, t)$,
which represent the grain mass distribution weighted for the grain mass and volume, respectively
(the latter can also be regarded as the volume distribution function).
The integration of $\rho(m,\, t)$ for $m$ is related to the dust-to-gas ratio, $\mathcal{D}$, as
\begin{align}
\mu_\mathrm{H}m_\mathrm{H}n_\mathrm{H}\mathcal{D}=\int_0^\infty\rho (m,\, t)\,\mathrm{d}m,
\label{eq:dg}
\end{align}
where $\mu_\mathrm{H}=1.4$ is the gas mass per hydrogen, $m_\mathrm{H}$ is
the hydrogen atom mass, and $n_\mathrm{H}$ is the hydrogen
number density.

The Smolchowski equation contains the collision kernel, which is the product of the
collision cross-section and the relative speed of colliding pair.
The collision kernel for colliding grains with masses $m_1$ and $m_2$ is denoted
as $K_{m_1,m_2}$, which is evaluated in Section \ref{subsec:kernel}.

It is straightforward to derive the following general equations applicable
for both coagulation and shattering based on equations (6) and (7) of O09:
\begin{align}
\lefteqn{\frac{\upartial\rho (m,\, t)}{\upartial t} = -m\rho (m,\, t)\int_0^\infty
\frac{K_{m,m_1}}{mm_1}\rho (m_1,\, t)\mathrm{d}m_1}\nonumber\\
& +\int_0^\infty\int_0^\infty\frac{K_{m_1,m_2}}{m_1m_2}\rho (m_1,\, t)\rho(m_2,\, t)
m\bar{\theta} (m;\, m_1,\, m_2)\mathrm{d}m_1\mathrm{d}m_2,\label{eq:rho}
\end{align}
\begin{align}
\lefteqn{\frac{\upartial\psi (m,\, t)}{\upartial t} = -\bar{V}(m,\, t)\psi (m,\, t)\int_0^\infty
\frac{K_{m,m_1}}{\bar{V}(m,\, t)\bar{V}(m_1,\, t)}\psi (m_1,\, t)\mathrm{d}m_1}\nonumber\\
& +\int_0^\infty\int_0^\infty\frac{K_{m_1,m_2}}{\bar{V}(m_1)\bar{V}(m_2)}
\psi (m_1,\, t)\psi(m_2,\, t)
(V_{1+2})_{m_1,m_2}^m\bar{\theta} (m;\, m_1,\, m_2)\mathrm{d}m_1\mathrm{d}m_2,
\label{eq:psi}
\end{align}
where $\tilde{\theta}(m;\, m_1,\, m_2)$ describes the distribution function of
grains produced from $m_1$ in the collision between grains with masses $m_1$ and $m_2$, and
$(V_{1+2})_{m_1,m_2}^m$ is the volume of the newly produced grain with mass $m$
in the above collision.
Since the expression is not exactly the same as that used by O09 and O12,
we explain the derivation of the above equations in Appendix \ref{app:derivation}.
In particular, in our formulation, we count the collisional products originating
from $m_1$ and $m_2$ separately. O09, in contrast, considered the collisional pair
$m_1$ and $m_2$ only once. This is why O09 has a factor 1/2 in front of the
second term on the
right-hand side in equations (\ref{eq:rho}) and  (\ref{eq:psi}).

We aim at solving equations (\ref{eq:rho}) and (\ref{eq:psi}) for
$\rho$ and $\psi$. Since $\rho /m=\psi /\bar{V}$ is always satisfied,
$\bar{V}$ is not an independent quantity. Thus, these two equations are closed
if we give $K$, $V_{1+2}$ and $\bar{\theta}$. In the following subsections, we explain how to
determine these three quantities.

In computing the grain size distribution, we discretize the entire grain radius range
($a=3\times 10^{-4}$--10 $\micron$) into 128
grid points with logarithmically equal spacing.
We set ${\rho}_\mathrm{d}(m,\, t)=0$ at the maximum and minimum grain radii for
the boundary conditions. We use the algorithm described in appendix B of
\citet{Hirashita:2019aa} to solve the discretized equations.

\subsection{Collision kernel}\label{subsec:kernel}

The collision kernel is determined by the product of the grain cross-section and
the relative velocity. The cross-section ($\sigma_{1,2}$) in the collision of grains
with masses $m_1$ and $m_2$, referred to as grain 1 and 2, respectively,
is estimated
as $\sigma_{1,2}=\upi (a_\mathrm{ch1}+a_\mathrm{ch2})^2$,
where the characteristic radii of grains 1 and 2 are
$a_\mathrm{ch1}$ and $a_\mathrm{ch2}$, respectively
(see the beginning of Section \ref{subsec:setup} for the definition of the characteristic radius;
i.e.\ $a_\mathrm{ch}=a_m\phi_m^{-1/3}$).

The motion of dust grains is assumed to be induced by turbulence
\citep[e.g.][]{Kusaka:1970aa,Volk:1980aa}. We basically adopt the same simple model,
originally taken from \citet{Ormel:2009aa}, for
the grain velocity $v_\mathrm{gr}$ as adopted in our previous papers:
\begin{align}
v_\mathrm{gr}(m) &= 1.1\mathcal{M}^{3/2}\phi_m^{1/3}\left(
\frac{a_m}{0.1~\micron}\right)^{1/2}\left(\frac{T_\mathrm{gas}}{10^4~\mathrm{K}}\right)^{1/4}
\left(\frac{n_\mathrm{H}}{1~\mathrm{cm}^{-3}}
\right)^{-1/4}\nonumber\\
&\times \left(\frac{s}{3.5~\mathrm{g~cm}^{-3}}\right)^{1/2}~\mathrm{km~s}^{-1},
\label{eq:vel}
\end{align}
where $\mathcal{M}$ is the Mach number of the largest-eddy velocity, and
$T_\mathrm{gas}$ is the gas temperature.
This is similar to equation (18) of \citet{Hirashita:2019aa}, but modified to include the
filling factor $\phi_m$ (see appendix A of \citealt{Ormel:2009aa}).
The derivation of the above equation does not strictly hold for highly supersonic regime,
but we practically use $\mathcal{M}$ here as an adjusting parameter for the grain
velocity. Under fixed ISM conditions and grain material properties, the velocity is reduced to
a function of $m$.
Since the filling factor $\phi_m$ changes as a function of time,
$v_\mathrm{gr}(m)$ also varies with time.
In considering the collision rate between grains 1 and 2 with
$v_\mathrm{gr}(m_1)=v_1$ and $v_\mathrm{gr}(m_2)=v_2$,
we estimate the relative velocity $v_{1,2}$ by
\begin{align}
v_{1,2}=\sqrt{v_1^2+v_2^2-2v_1v_2\mu_{1,2}\,}\,,\label{eq:rel_vel}
\end{align}
where $\mu =\cos\theta$ ($\theta$ is an angle between the two grain velocities)
is randomly chosen between $-1$ and 1 in every calculation of the collision kernel
\citep{Hirashita:2013aa}.

Using the above quantities, the collision kernel is estimated for the collision between
grains 1 and 2 as
\begin{align}
K_{m_1,m_2}=\sigma_{1,2}v_{1,2}.
\end{align}
Note that the collision kernel depends on the filling factor of the larger grain as
$\propto\phi_m^{-1/3}$. Thus, the grain--grain collision rate is enhanced if
the grains become more porous (fluffy).

\subsection{Treatment of grains produced by collisions}

The treatment of collisional products is the key for the evolution of porosity.
The porosity is strongly affected by the production of vacuum volume in a grain
after grain--grain sticking (coagulation) and the compaction associated with
compression in grain--grain collisions. However, these processes are strongly
nonlinear and dependent on the actual grain shapes. Moreover, some assumptions in
O09 and O12 are not applicable to interstellar dust. In particular, their formulation is
based on single-sized ($\sim 0.1~\micron$) monomers, since they are interested in protoplanetary discs,
where grains grow into much larger sizes than interstellar grains. In the
ISM, in contrast, we also consider the evolution of grains much smaller than the
`typical' grain size ($\sim 0.1~\micron$). To make the problem more complicated,
shattering also occurs, producing a large number of small grains.
Because of the above difficulty and the difference from O09 and O12, we
take the following approach that simplifies the treatment of porosity evolution
but still preserves the essence of physical processes regarding grain--grain sticking and
compression.

\subsubsection{Coagulation}\label{subsubsec:coag}

For coagulation, two characteristic energies are important: impact (kinetic) energy
and the rolling energy.
The impact energy ($E_\mathrm{imp}$) in the collision between grains 1 and 2 is estimated as
\begin{align}
E_\mathrm{imp}=\frac{1}{2}\frac{m_1m_2}{m_1+m_2}v_\mathrm{1,2}^2\, .
\end{align}
The rolling energy ($E_\mathrm{roll}$)  is defined as
the energy necessary for a monomer to roll over 90 degrees on the surface
of another monomer \citep{Dominik:1997aa}, and
is given by \citep{Wada:2007aa}
\begin{align}
E_\mathrm{roll}=12\pi^2\gamma R_{1,2}\xi_\mathrm{crit},\label{eq:E_roll}
\end{align}
where $\gamma$ is the surface energy per unit contact area, $R_{1,2}$ is the
reduced particle radius, and $\xi_\mathrm{crit}$ is the critical displacement of rolling.
The surface energy depends on the surface material:
$\gamma =25$, 75, and 100 erg cm$^{-2}$ for silicate (originally from quartz),
graphite, and water ice, respectively \citep{Dominik:1997aa,Israelachvili:1992aa,Wada:2007aa}.
The critical displacement $\xi_\mathrm{crit}$ lies broadly in the range of
$\sim$2--30 \AA\ \citep{Dominik:1997aa,Heim:1999aa,Wada:2007aa}.
As shown by \citet{Kimura:2015aa} and \citet{Steinpilz:2019aa},
the surface energy of dry silica could be larger than the above values.
This implies that the above surface energy for silicate is
underestimated.
Since $\gamma$ and $\xi_\mathrm{crit}$ are degenerate, we fix $\gamma$ to an
intermediate value (75 erg cm$^{-2}$) and vary $\xi_\mathrm{crit}$.
We choose a fiducial value $\xi_\mathrm{crit}=10$ \AA\ unless otherwise stated
but we later examine a case where $\xi_\mathrm{crit}$ is varied.

The reduced radius, $R_{1,2}$, is difficult
to determine rigidly for interstellar dust, since the typical monomer size is not obvious.
Thus, we simply assume that
\begin{align}
\frac{1}{R_{1,2}}=\frac{1}{a_{m_1}}+\frac{1}{a_{m_2}}.
\end{align}
This equation gives a reasonable estimate for $R_{1,2}$ for compact grains.
As shown later, small grains tend to be compact. Since the reduced radius is dominated by
the smaller grain, the above estimate could be justified.
Large (submicron/micron) grains are also affected by compaction, so that the above reduced radius
could be applied for collisions between large grains.
However, the above reduced mass may be overestimated in
collisions between porous grains. Grains with intermediate radii ($a\sim 0.01$--0.1~$\micron$)
tend to be porous. Overestimation of $R_\mathrm{1,2}$ leads to
less efficient compaction. On the other hand, the uncertainty in $R_{1,2}$ could be
absorbed by the variation of $\xi_\mathrm{crit}$ and $n_\mathrm{c}$
($n_\mathrm{c}$ is introduced below).
Thus, we vary $\xi_\mathrm{crit}$ and $n_\mathrm{c}$ to effectively investigate the uncertainties
caused by collisions between
intermediate-sized grains.

We now estimate the volume of the coagulated grain.
We do not directly adopt O12's formulation (their equation 15; see also \citealt{Suyama:2012aa})
because, as mentioned above, the monomer size is not well determined for
interstellar dust. Nevertheless, we apply the following essential points
\citep{Dominik:1997aa}:
(i) The physical outcome of the collision is characterized by the ratio between
$E_\mathrm{imp}$ and $E_\mathrm{roll}$.
(ii) If $E_\mathrm{imp}\ll E_\mathrm{roll}$, the grains stick without modifying their
structures (hit-and-stick collisions). (iii) If $E_\mathrm{imp}\sim E_\mathrm{roll}$,
the newly added void tends to be compressed because the contact points of monomers
are moved. (iii) If $E_\mathrm{imp}\gtrsim n_\mathrm{c}E_\mathrm{roll}$,
where $n_\mathrm{c}$ is the number
of contact points, significant compaction occurs.
Since we do not trace each monomer, $n_\mathrm{c}$ is uncertain.
Moreover, not all contact points are equally important in compaction.
Thus, we treat $n_\mathrm{c}$ as a free parameter, and regard it as
the number of major contacts whose movement contributes to compaction significantly.
Since $n_\mathrm{c}$ always appears in the form of product $n_\mathrm{c}E_\mathrm{roll}$,
$n_\mathrm{c}$ is degenerate with $E_\mathrm{roll}$ as mentioned above.
Considering the above three points, (i)--(iii), we estimate the
volume ($V_{1+2}^0$) of coagulated product in the collision between grains with volumes
$V_1$ and $V_2$ (assuming that $V_1\geq V_2$; if $V_1<V_2$, we exchange grains 1 and 2, so that
the following formulation holds) as
\begin{align}
V_{1+2}^0 &= V_1+V_2+V_\mathrm{void}\exp\left[-E_\mathrm{imp}/(3bE_\mathrm{roll})\right]
\nonumber\\
&+(V_\mathrm{2,comp}-2V_2)\left\{1-\exp\left[ -E_\mathrm{imp}/(n_\mathrm{c}E_\mathrm{roll})\right]\right\} ,
\label{eq:v120}
\end{align}
where $V_\mathrm{void}$ is the volume of the void newly created in the collision,
$b=0.15$ is an adjusting parameter \citep{Wada:2008aa}, and
$V_\mathrm{2,comp}\equiv m_2/s$ (the volume with perfect compaction).
The void volume is estimated as (O12)
\begin{align}
V_\mathrm{void}=\min\left[ 0.99-1.03\ln\left(\frac{2}{V_1/V_2+1}\right),\, 6.94\right] V_2.
\end{align}
Equation (\ref{eq:v120}) satisfies the above conditions (i)--(iii):
$V_{1+2}^0\simeq V_1+V_2+V_\mathrm{void}$
for $E_\mathrm{imp}\ll E_\mathrm{roll}$. The newly created volume ($V_\mathrm{void}$) is compressed
if $E_\mathrm{imp}\gtrsim E_\mathrm{roll}$. If $E_\mathrm{imp}$ becomes comparable
to or higher than $n_\mathrm{c}E_\mathrm{roll}$, the grain is compressed further (note that
$V_\mathrm{2,comp}-2V_2\leq 0$). If $E_\mathrm{imp}\gg n_\mathrm{c}E_\mathrm{roll}$,
$V_{1+2}^0=(V_1-V_2)+V_\mathrm{2,comp}$ (note that $V_1\geq V_2$), which means that
$V_2$ becomes compact while $V_1$ is compressed by the equivalent volume to
$V_2$. According to \citet{Wada:2013aa},
in a collision of grains with different sizes, the larger grain is not entirely compressed.
We assume that the compressed volume in the larger grain is determined by the volume of
the smaller grain. In the above expression, $V_{1+2}^0$ can be smaller than
$V_\mathrm{1,comp}+V_\mathrm{2,comp}$, where $V_\mathrm{1,comp}=m_1/s$
(the volume with perfect compaction). Thus, we finally adopt the following expression for
the coagulated volume, $V_{1+2}$:
\begin{align}
V_{1+2}=
\begin{cases}
V_{1+2}^0 & \mbox{if $V_{1+2}^0>V_\mathrm{1,comp}+V_\mathrm{2,comp}$,} \\
(1+\epsilon_V)(V_\mathrm{1,comp}+V_\mathrm{2,comp}) & \mbox{otherwise,}
\end{cases}
\label{eq:V12}
\end{align}
where $\epsilon_{V}\geq 0$ is a parameter reflecting the fact that
perfect compaction is difficult \citep{Suyama:2012aa,Wada:2013aa}.
We use this volume for $(V_{1+2})_{m_1,m_2}^m$ in equation (\ref{eq:psi}).

The mass distribution of collisional products is described by
(see equation 26 of \citealt{Hirashita:2019aa})
\begin{align}
m\bar{\theta}(m;\, m_1,\, m_2)={m_1}\delta [m-(m_1+m_2)],
\end{align}
where $\delta (\cdot )$ is Dirac's delta function.
As mentioned above and in Appendix \ref{app:derivation}, we separately
treat the collision product originating from $m_1$ and that from $m_2$ (i.e.\ we count the
same collision twice).

\subsubsection{Shattering}\label{subsubsec:shat}

There has not been any formulation for the evolution of
porosity by shattering in the ISM. Shattering produces a lot of fragments, and sometimes leaves
a remnant if the original grain is much larger than the colliding partner.
We expect that shattered fragments have small
porosities because weakly bound parts become unbound in disruptive
collisions. Thus, we assume that the collisional fragments are compact.
For the remnant, it is probable compaction occurs;
however, in our formulation, the remnant remains only if a grain collides with
a much smaller grain: in this case, it is not clear if the entire remnant is efficiently
pressed. Thus, we simply assume that the shattered remnant has the same porosity
as the original grain. We examine separately a case where we take compaction of
the remnants into account.

The microscopic processes for shattering of porous grains in high-velocity
($\gtrsim 1$ km s$^{-1}$) collisions are not well known.
Moreover, in our formulation, we have to treat collisions of
both porous and compact grains. Thus, we use our previous formulation in
\citet{Hirashita:2019aa}, which is appropriate for compact grains.
We also expect that the resulting grain size
distribution is not sensitive to detailed assumptions on the fragment size
distribution \citep{Hirashita:2013ab}. We explain the treatment of shattered fragments
in what follows.

We consider a collision of two dust grains with masses $m_1$
and $m_2$ (grains 1 and 2) based on \citet{Kobayashi:2010aa}'s model.
The ejected mass ($m_\mathrm{ej}$) from grain 1 is estimated as
\begin{align}
m_\mathrm{ej}=\frac{\varphi}{1+\varphi}m_1,
\end{align}
with
\begin{align}
\varphi\equiv\frac{E_\mathrm{imp}}{m_1Q_\mathrm{D}^\star},
\end{align}
where $Q_\mathrm{D}^\star$
is the specific impact energy required to disrupt half of the mass ($m_1/2$).
We adopt $Q_\mathrm{D}^\star =8.9\times 10^9$ erg g$^{-1}$ following HM20.
The ejected mass is distributed into fragments, for which
we assume a power-law size distribution with an index of
$\alpha_\mathrm{f}=3.3$ \citep{Jones:1996aa}.
{This index is translated into that of $\bar{\theta}$ as
$\bar{\theta}\propto m^{-(\alpha_\mathrm{f}+2)/3}$.}
The maximum and minimum masses of the fragments
are assumed to be
$m_\mathrm{f,max}=0.02m_\mathrm{ej}$ and
$m_\mathrm{f,min}=10^{-6}m_\mathrm{f,max}$, respectively \citep{Guillet:2011aa}.
We adopt the following mass distribution function of collisional product from
grain 1 in the collision with grain 2
as
\begin{align}
m\bar{\theta}(m,\, m_1,\, m_2) &=
\frac{(4-\alpha_\mathrm{f})m_\mathrm{ej}m^{(-\alpha_\mathrm{f}+1)/3}}{3\left[
m_\mathrm{f,max}^\frac{4-\alpha_\mathrm{f}}{3}-
m_\mathrm{f,min}^\frac{4-\alpha_\mathrm{f}}{3}\right]}\,
\Phi (m;\, m_\mathrm{f,min},\, m_\mathrm{f,max})\nonumber\\
&+m\delta (m-m_1+m_\mathrm{ej}),\label{eq:frag}
\end{align}
where
$\Phi (m;\, m_\mathrm{f,min},\, m_\mathrm{f,max})=1$ if
$m_\mathrm{f,min}\leq m\leq m_\mathrm{f,max}$, and 0 otherwise.
Grains which become smaller than the minimum grain size
($a_m=3\times 10^{-4}~\micron$)
are removed.
For the volumes of shattered products, we adopt
\begin{align}
V_{1+2}=
\begin{cases}
\bar{V}(m_1)/(1+\varphi ) & \mbox{if $m=m_1-m_\mathrm{ej}=m_1/(1+\varphi )$,}\\
m/s & \mbox{otherwise.}
\end{cases}
\label{eq:V12_shat}
\end{align}
The first case of this expression indicates that the porosity (or filling factor)
of the remnant is the same as that of the original grain. The second case
means that the fragments are compact (that is, the volumes are simply estimated by
the mass divided by the material density).
We use equation (\ref{eq:V12_shat}) for $(V_{1+2})_{m_1,m_2}^m$ in equation (\ref{eq:psi}).

\subsection{Calculation of extinction curves}\label{subsec:ext_method}

To investigate the effects on observed dust properties, we also calculate extinction curves.
{We examine two representative grain materials: silicate and carbonaceous species, and
investigate how the porosity evolution driven by coagulation and shattering
affects the extinction properties of these materials.
For silicate,
we use the optical constants of astronomical silicate adopted
by \citet{Weingartner:2001aa}. For carbonaceous dust, we adopt amorphous carbon (amC)
from `ACAR' in \citet{Zubko:1996aa}. Graphite is another possible carbonaceous material,
and we confirmed that it produces similar results to amC except at wavelengths
where the 2175 \AA\ feature is prominent. We also found that
the wavelength where
the feature peaks shifts with the porosity, but such a shift is not observed in the
Milky Way extinction curves. Thus, special care should be taken of the modelling of
the 2175 \AA\ feature, e.g.\ by treating the 2175 \AA\ carrier such as graphite and PAHs
\citep[e.g.][]{Li:2001aa} as a component
separated from porous grain species \citep{Voshchinnikov:2006aa},
which is out of the scope of this paper.
Thus, we adopt amC in this paper, noting that, as mentioned above, amC and graphite
produce similar results at wavelengths not affected by the 2175 \AA\ feature.}

The optical properties of dust is calculated using the effective medium theory
(EMT), which averages the dielectric permittivity using a mixing rule.
We adopt the Bruggeman mixing rule.
This mixing rule as well as the Garnett mixing rule gives reasonable extinctions as long as
the grains do not have substructures (e.g.\ monomers) larger than the wavelength
\citep{Voshchinnikov:2005aa}.
As shown later, the porous grains are formed by coagulation of grains
smaller than $\sim 0.1~\micron$ and we are interested in wavelengths longer than
0.1 $\micron$. Thus, the above condition is satisfied in this paper.
Even if we model the inhomogeneity using
multi-layered spheres, the difference in the extinction is expected to be less than
$\sim 10$ per cent \citep{Voshchinnikov:2005aa,Shen:2008aa}.
Using the Bruggeman mixing rule, the averaged dielectric permittivity $\bar{\varepsilon}$
is obtained from
\begin{equation}
(1-\phi_m) \frac{1 - \bar{\varepsilon}}{1 + 2 \bar{\varepsilon}} + 
\phi_m \frac{\varepsilon_2 - \bar{\varepsilon}}{\varepsilon_2 + 2 \bar{\varepsilon}} = 0,
\end{equation}
where $\varepsilon_2$ is the dielectric permittivity of the material
(note that the first material is assumed to be vacuum so
$\varepsilon_1=1$).\footnote{{We adopt Gaussian-cgs units.}}
We assume that each dust grain is a sphere with radius $a_\mathrm{ch}$ and refractive index
$\bar{m}=\sqrt{\bar{\varepsilon}}$.
The cross-section $C_{\mathrm{ext},m}$, which is a function of $m$ under a given $\phi_m$ at
each epoch,
is calculated by the Mie theory \citep[][]{Bohren:1983aa}.

The extinction at wavelength $\lambda $ in units of magnitude ($A_{\lambda }$) can be
calculated using the grain size distribution as
\begin{align} 
A_{\lambda}=(2.5\log_{10} \mathrm{e})L\displaystyle\int_{0}^{\infty}
\tilde{n}(m)\,C_{\mathrm{ext},m}\,\mathrm{d}m,
\end{align}
where $L$ is the path length.
We present the extinction in the following two ways: $A_\lambda/N_\mathrm{H}$ and
$A_\lambda /A_V$ (the $V$ band wavelength corresponds to
$\lambda ^{-1}=1.8\,\mu {\rm m}^{-1}$ and $N_\mathrm{H}=n_\mathrm{H}L$ is the column density of
hydrogen nuclei). The first quantity indicates the extinction per hydrogen (thus,
it is proportional to the dust abundance), while
the second is useful to show the shape of extinction curve. In both quantities,
the path length $L$ is canceled out.

\subsection{Parameter settings}\label{subsec:param}

Coagulation and shattering are governed by the same form of equation
(equations \ref{eq:rho} and \ref{eq:psi}). These two processes occur in different ISM phases.
We assume that coagulation occurs in the dense clouds of which the physical conditions are described by
$n_\mathrm{H}=10^3$ cm$^{-3}$ and $T_\mathrm{gas}=10$ K (typical of molecular clouds);
and that
shattering takes place in the diffuse ISM characterized by
$n_\mathrm{H}=0.3$ cm$^{-3}$ and $T_\mathrm{gas}=10^4$ K.
These values are similar to the ones adopted for coagulation and shattering
by \citet{Hirashita:2009ab}.
The density affects the grain collision time-scale, which scales with
$(n_\mathrm{H}v_\mathrm{gr})^{-1}\propto n_\mathrm{H}^{-3/4}$ (see equation \ref{eq:vel});
that is, the adopted duration for shattering/coagulation is degenerate with the
gas density. Therefore, we fix the above physical conditions for the gas and examine the
grain size distributions at various times (i.e.\ for various durations of the process),
keeping in mind that the same value of $n_\mathrm{H}^{3/4}t$ gives similar results.

For the normalization of the grain velocities adjusted by $\mathcal{M}$ in equation (\ref{eq:vel}),
we apply $\mathcal{M}=3$ for shattering and $\mathcal{M}=1$ for coagulation
(HM20). These values of $\mathcal{M}$ are adopted to obtain a similar level of
grain velocities to those calculated for magnetized turbulence by \citet{Yan:2004aa}.

We not only treat shattering and coagulation separately, but also examine some cases
where these two processes occur at the same time. The coexistence of shattering and
coagulation is a reasonable approximation for a volume of the ISM wide enough
to sample a statistically significant amount of both ISM  phases and/or for a time much longer than
the mixing time-scale of the two ISM phases ($\sim 10^7$ yr; e.g.\ \citealt{McKee:1989aa}).
Since coagulation and shattering occur
in the media with different
densities, it is convenient to define the grain mass and volume distributions per hydrogen number density
as $\tilde{\rho}\equiv\rho /n_\mathrm{H}$ and $\tilde{\psi}\equiv\psi /n_\mathrm{H}$, respectively.
In this case, we calculate the evolution of $\tilde{\rho}$ and $\tilde{\psi}$ by
\begin{align}
\left.\frac{\upartial\tilde{\rho}(m,\, t)}{\partial t}\right|_\mathrm{tot} &=
\eta_\mathrm{dense}\left.\frac{\upartial\tilde{\rho}(m,\, t)}{\partial t}\right|_\mathrm{coag}
+(1-\eta_\mathrm{dense})\left.\frac{\upartial\tilde{\rho}(m,\, t)}{\partial t}\right|_\mathrm{shat},
\label{eq:rho_mix}\\
\left.\frac{\upartial\tilde{\psi}(m,\, t)}{\partial t}\right|_\mathrm{tot} &=
\eta_\mathrm{dense}\left.\frac{\upartial\tilde{\psi}(m,\, t)}{\partial t}\right|_\mathrm{coag}
+(1-\eta_\mathrm{dense})\left.\frac{\upartial\tilde{\psi}(m,\, t)}{\partial t}\right|_\mathrm{shat},
\label{eq:psi_mix}
\end{align}
where $\eta_\mathrm{dense}$, which is treated as a free parameter, is
the dense gas fraction determining the weight of each process, the subscript
`tot' indicates the total changing rates of $\tilde{\rho}$ and $\tilde{\psi}$, and the subscripts
`coag' and `shat'
mean the changes caused by coagulation and shattering, respectively.
For the two terms on the right-hand side of the above equations, we apply
equations (\ref{eq:rho}) and (\ref{eq:psi}).
The change by coagulation and that by shattering are mixed with a ratio of
$\eta_\mathrm{dense}:(1-\eta_\mathrm{dense})$; in other words, $\eta_\mathrm{dense}$
determines the fraction of time dust spends in the dense ISM.
In calculating shattering and coagulation, we use $\rho$ and $\psi$, but when we sum up
the contributions from coagulation and shattering, we divide them by $n_\mathrm{H}$ and
obtain $\tilde{\rho}$ and $\tilde{\psi}$.
Here we implicitly neglect gas which hosts neither coagulation nor shattering.
Existence of such a gas component effectively lowers the efficiencies of
both shattering and coagulation
(or makes the time-scales of these two processes longer)
equally and does not affect the relative roles of these two processes.

\subsubsection{Important parameters specific for coagulation}

As mentioned in Section \ref{subsubsec:coag}, because of the degeneracy between $\gamma$
and $\zeta_\mathrm{crit}$ (equation \ref{eq:E_roll}),
we fix $\gamma =75$ erg cm$^{-2}$ and vary $\zeta_\mathrm{crit}=2$--30~\AA.
We here adopt $\zeta_\mathrm{crit}=10$ \AA\ as a fiducial value.
Since the grain--grain collision velocities are typically less than 50 m s$^{-1}$ in the dense ISM,
we assume for coagulation that the grains always stick when they collide \citep{Wada:2013aa}.

The parameter that regulates the maximum compaction, $\epsilon_V$
(equation \ref{eq:V12}) is also an unknown parameter. As shown later,
this parameter basically determines the filling factor of grains larger than submicron.
Although this parameter is uncertain, we argue later that $\epsilon_V\lesssim 1$.
If $\epsilon_V$ is larger than 1, it imposes an artificial fluffiness for submicron grains
as we discuss in Section \ref{subsec:coag}.
We adopt $\epsilon_V=0.5$ for the fiducial value but examine the variation of
$\epsilon_V=0$--1 separately.

The number of contact points, $n_\mathrm{c}$, is also unknown since we do not trace
each monomer. We interpret this parameter as the number of contacting points whose
motion significantly reduces the porosity. We assume $n_\mathrm{c}=30$ unless otherwise stated.
We also examine the effect of $n_\mathrm{c}$ by changing its value.

\subsubsection{Important parameters specific for shattering}\label{subsubsec:param_shat}

In the above, we assumed for shattering that the fragments are compact while the remnant
has the same filling factor as the original grain (equation~\ref{eq:V12_shat}).
However, compaction could occur for the remnants. Here, for an experimental purpose,
we consider a model in which
the comparable volume to the ejecta suffers compaction. Noting that the ejecta have a
fraction $\varphi /(1+\varphi )$ of the original grain, the compaction of the volume corresponding to
that fraction can be written as
\begin{multline}
V_{1+2} = \max\left[\frac{1-\varphi}{1+\varphi}V(m_1) + \frac{\varphi}{1+\varphi}\frac{m_1}{s},\,
\frac{m_1}{(1+\varphi )s}\right]\\
\text{if $m=m_1/(1+\varphi )$}.\label{eq:compact_rem}
\end{multline}
The first case in the max function describes the replacement of the volume
$[\varphi /(1+\varphi )]V(m_1)$ (the original total volume of the ejecta) with
$[\varphi /(1+\varphi )]m_1/s$ (the corresponding compact volume).
However, this breaks down for large $\varphi$, and $V_{1+2}$ could even become negative.
Thus, we set the second case in the max function, which describes the fully compact remnant.
By default, we use equation~(\ref{eq:V12_shat}), but when we include compaction of remnants, 
we use equation (\ref{eq:compact_rem}) instead of the first condition in equation (\ref{eq:V12_shat}).

\subsubsection{Initial condition}

Coagulation and shattering basically conserve the total grain mass.
Strictly speaking, because we set the upper and lower boundaries for the grain radii
($a_\mathrm{min}=3$~\AA\ and $a_\mathrm{max}=10~\micron$, respectively; Section \ref{subsec:setup})
and remove all grains coagulated or shattered beyond the
boundaries, the total grain mass is not strictly conserved. Except for this effect,
our algorithm guarantees the conservation of the total grain mass
(see appendix B of \citealt{Hirashita:2019aa}). We adopt a power-law grain size distribution
similar to the MRN distribution:
$n_\mathrm{init}(a_m)\propto a_m^{-p}$ ($p=3.5$) with
the lower and upper bounds of grain radii being $a_\mathrm{min,ini}=0.001~\micron$ and
$a_\mathrm{max,ini}=0.25~\micron$, respectively.\footnote{{Note that
$a_\mathrm{min/max,ini}$ and $a_\mathrm{min/max}$ are different with the latter chosen
to be similar to the constraint from MRN.
We also confirmed that
the results below are insensitive to $a_\mathrm{min,ini}$
as long as $a_\mathrm{min,ini}\lesssim 0.01~\micron$.
By the nature of coagulation, grains with $a<a_\mathrm{min,ini}$ do not appear in the pure coagulation
cases below.}}
This initial grain size distribution is related to the above grain mass distribution as
$n_\mathrm{init}(a_m)\,\mathrm{d}a_m=\tilde{n}(m,\,t=0)\,\mathrm{d}m$.
With the above power-law grain size distribution, we obtain, recalling that
$\rho (m,\, t)=m\tilde{n}(m,\, t)$,
\begin{align}
\rho (m,\, t=0)=
\frac{(4-p)\mu_\mathrm{H}m_\mathrm{H}n_\mathrm{H}\mathcal{D}}
{3\left[ m_\mathrm{max,ini}^{(4-p)/3}-m_\mathrm{min,ini}^{(4-p)/3}\right]}
m^{(-p+1)/3}
\end{align}
for $m_\mathrm{min,ini}<m<m_\mathrm{max,ini}$,
where $m_\mathrm{max,ini}=4\pi a_\mathrm{max,ini}^3s/3$ and
$m_\mathrm{min,ini}=4\pi a_\mathrm{min,ini}^3s/3$.
We used equation (\ref{eq:dg}) for normalization.
Thus, if we give $\mathcal{D}$, we set the initial condition.
We adopt $\mathcal{D}=0.01$ as a typical dust-to-gas ratio in
the Milly Way, but remind the reader that the time-scales of
coagulation and shattering scale with $\mathcal{D}^{-1}$.
The initial filling factor is assumed to be the same (unity unless otherwise stated)
for all grains because we do not know the filling factor in advance.

Note that the initial condition here is not meant to represent the initial
grain size distribution in a galaxy. It is aimed at setting a `standard' grain size distribution,
based on which we investigate the effect of coagulation and shattering.
Therefore, the MRN grain size distribution, which roughly reproduces the dust
properties (e.g.\ extinction curves) in the Milky Way is simply used as a starting point.
This initial condition also serves to understand how the extinction curves should be modified by
coagulation and shattering. This approach is similar to the one taken by
\citet{Hirashita:2009ab} for compact grains.

\section{Results}\label{sec:result}

We show the resulting grain size distributions, filling factors and extinction curves
in this section. We first examine coagulation and shattering separately to clarify the effects of
each process (these cases are referred to pure coagulation and pure shattering).
After that, we clarify the combined effects arising from the ISM phase exchange
by including both processes simultaneously (Section \ref{subsec:param}).

In the following, the grain size distribution is always shown in the form of
$a_m^4n(a_m)/n_\mathrm{H}$
(the variable $t$ in $n$ is omitted), where the grain size distribution $n(a_m)$ is defined as
$n(a_m)\,\mathrm{d}a_m=\tilde{n}(m)\,\mathrm{d}m$ [recall that $m=(4\upi /3)a_m^3s$].
Since $\rho (m)\,\mathrm{d}m=m\tilde{n}(m)\,\mathrm{d}m\propto a_m^3n(a_m)\,\mathrm{d}a_m
\propto a_m^4n(a_m)\,\mathrm{d}\log a_m$, $a_m^4n(a)$ is proportional to the mass distribution
function per $\log a_m$. We further divide this quantity by $n_\mathrm{H}$ to
cancel out the overall density difference between the dense and diffuse ISM.
We refer to $a_m^4n(a_m)/n_\mathrm{H}$ also as the grain size distribution whenever
there is no risk of confusion.

\subsection{Pure coagulation}\label{subsec:coag}

\begin{figure}
\includegraphics[width=0.45\textwidth]{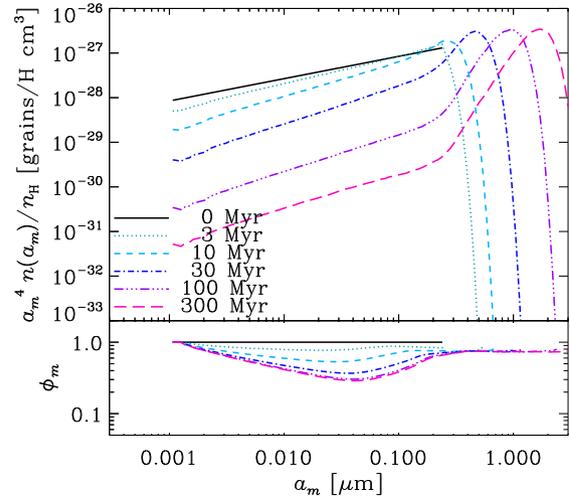}
\caption{Evolution of grain size distribution (upper window) and filling factor (lower window)
for the pure coagulation case.
The grain size distribution is multiplied by $a_m^4$ and divided by $n_\mathrm{H}$,
so that the resulting quantity is proportional to the grain mass abundance  per $\log a_m$
relative to the gas mass.
The solid, dotted, short-dashed, dot--dashed, triple-dot--dashed, and long-dashed lines show
the results
at $t=0$ (initial condition), 3, 10, 30, 100, and 300~Myr, respectively.
The time evolution of $\phi_m$ (filling factor) is also shown in the same line species as in the
upper window.
Note that we use the mass-equivalent grain radius $a_m$, while
the characteristic grain radius is obtained by $a_\mathrm{ch}=a_m\phi_m^{-1/3}$.
\label{fig:coag}}
\end{figure}

We examine the evolution of grain size distribution and filling factor by coagulation
in the dense ISM (the pure coagulation case). We show the results in Fig.~\ref{fig:coag}.
Coagulation continuously forms large grains, depleting small
grains. The bump in the grain size distribution at large grain radii
is prominent after coagulation takes place significantly.
The height of the bump is almost constant, reflecting the mass conservation
in coagulation. Micron-sized grains are formed on a time-scale of 100~Myr, which is
consistent with \citet[][see their figure 4a]{Hirashita:2014ab}.
Note that this time-scale scales with the assumed density roughly as
$\propto n_\mathrm{H}^{-3/4}$ as mentioned in Section \ref{subsec:param}
(recall that we adopt $n_\mathrm{H}=10^3$ cm$^{-3}$ for the dense ISM).

We also show the filling factor $\phi_m$ in Fig.\ \ref{fig:coag}. We observe that the
filling factor steadily decreases up to $t\sim 100$ Myr because coagulation creates
porosity. The decrease of filling factor is governed by coagulation of small grains,
which are, however, effectively depleted by coagulation.
As the abundance of small grains becomes lower, the
decrease of the filling factor is saturated.
For large ($a\gtrsim 0.1~\micron$) grains, the porosity is determined
by the assumption of maximum compaction regulated by $\epsilon_V$ in equation (\ref{eq:V12}).
The filling factor at large grain radii roughly approaches $\phi_m\sim 1/(1+\epsilon_V)
\simeq 0.67$ (recall that $\epsilon_V=0.5$). In fact, the value is slightly larger than 0.67,
because of
our treatment of equation (\ref{eq:V12}): There are still cases where
$V_\mathrm{1,comp}+V_\mathrm{2,comp}<V_{1+2}<
(1+\epsilon_V)(V_\mathrm{1,comp}+V_\mathrm{2,comp})$.
In this case, grains whose filling factors are between
$1/(1+\epsilon_V)$ and 1 form, contributing to raising the averaged $\phi_m$ above
$1/(1+\epsilon_V)$ at large grain radii.

\begin{figure}
\includegraphics[width=0.45\textwidth]{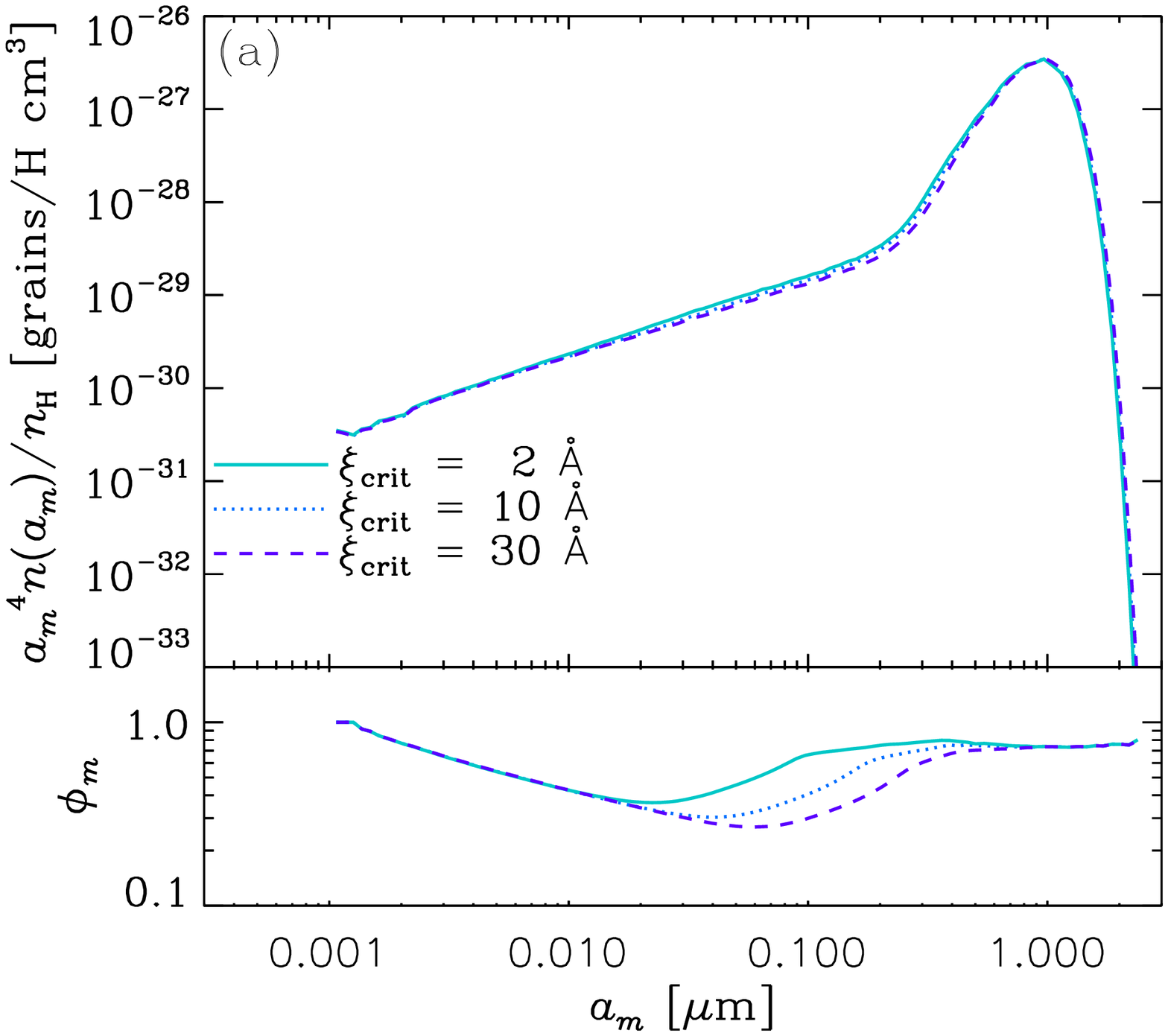}
\includegraphics[width=0.45\textwidth]{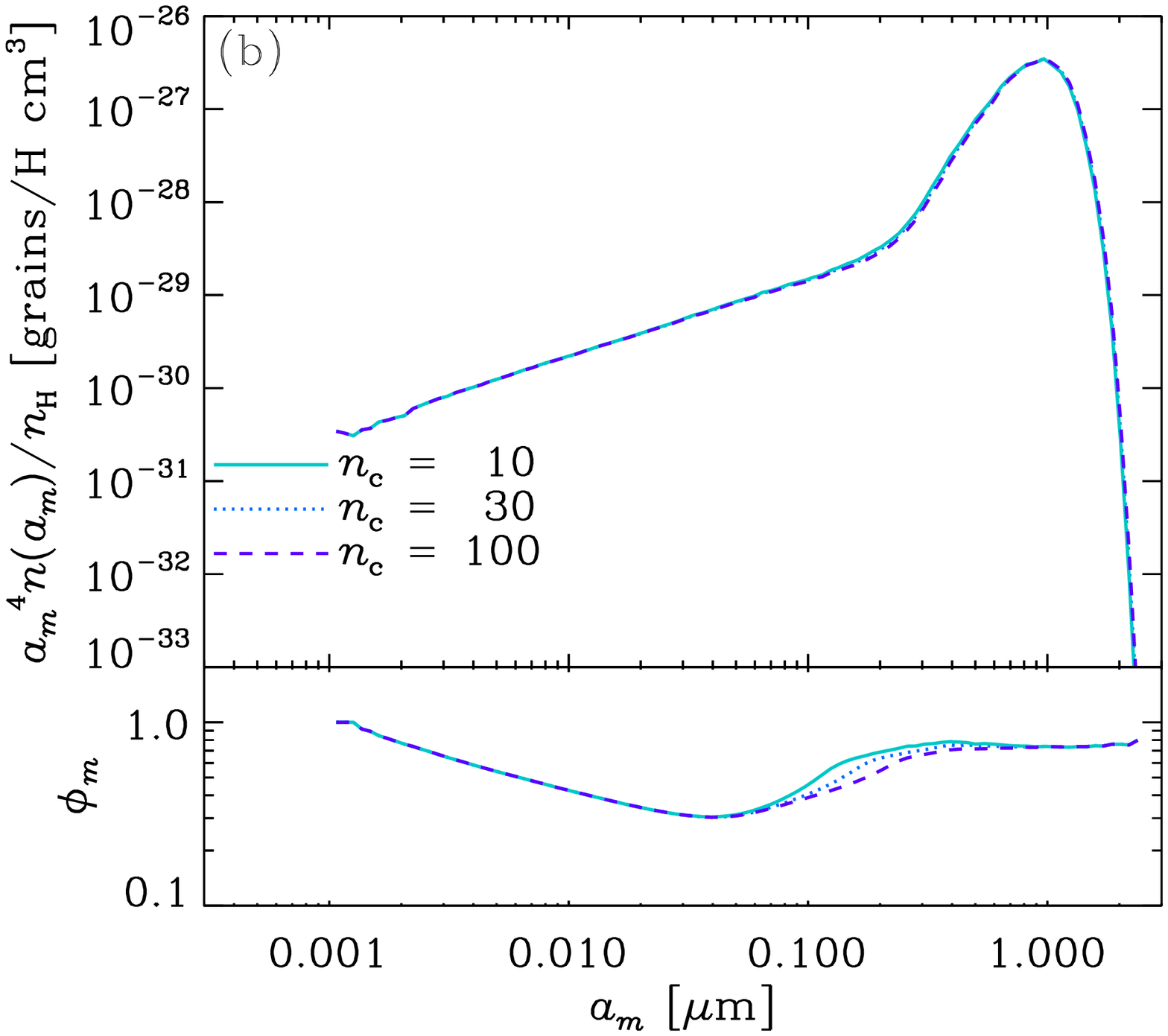}
\includegraphics[width=0.45\textwidth]{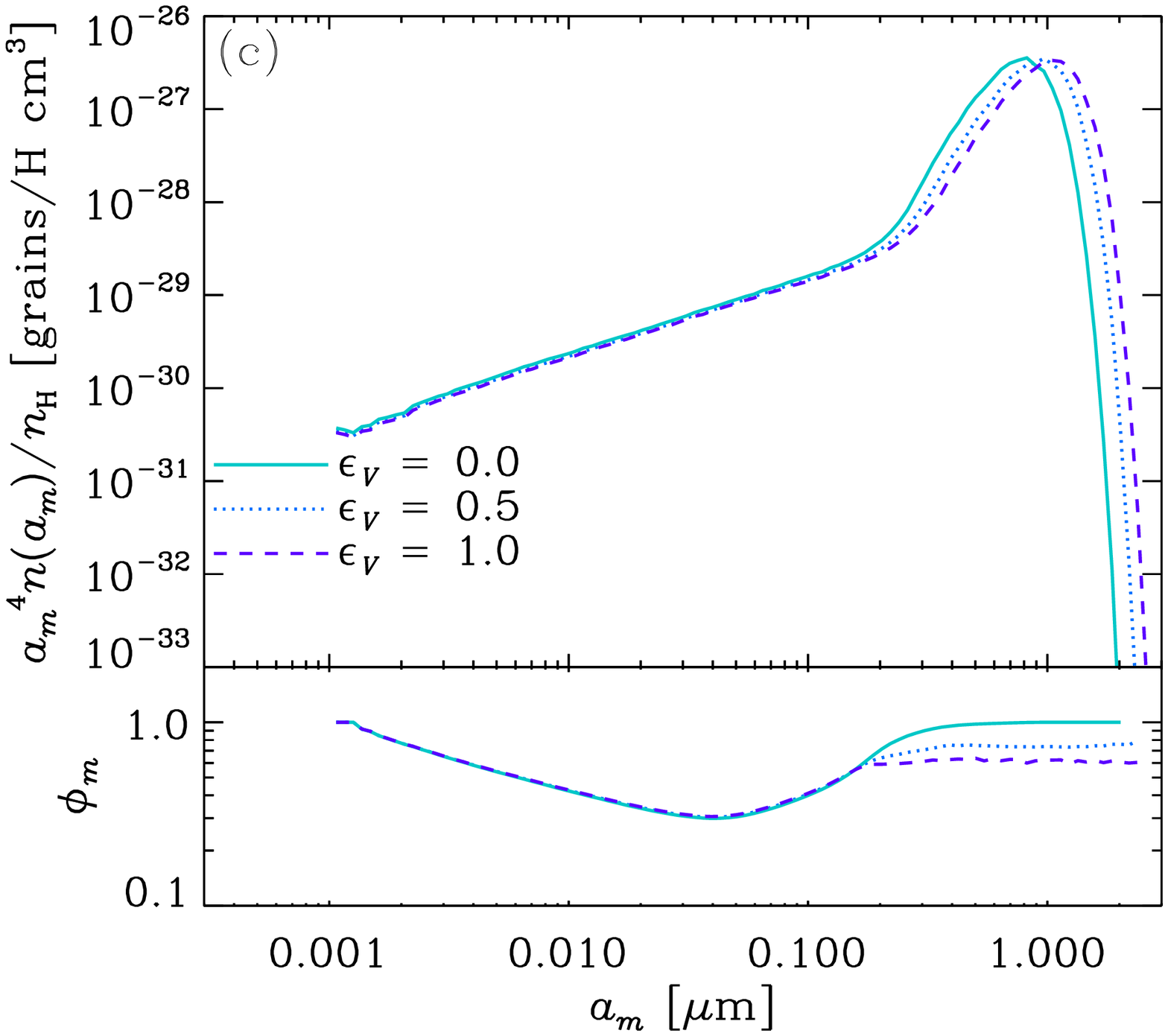}
\caption{Parameter dependence for the pure coagulation cases. We present the
grain size distributions (upper window) and the filling factors (lower window) at $t=100$~Myr.
(a) Dependence on $\xi_\mathrm{crit}$, which regulates $E_\mathrm{roll}$.
The solid, dotted, and dashed lines show the results for $\xi_\mathrm{crit}=2$, 10, and 30~\AA,
respectively. (b) Dependence on $n_\mathrm{c}$, which regulates compaction.
The solid, dotted, and dashed lines show the results for
$n_\mathrm{c}=10$, 30, and 100, respectively.
(c) Dependence on $\epsilon_V$, which regulates the maximum compaction in
high-velocity collisions (relevant for large grains). The solid, dotted, and dashed lines show
the results for $\epsilon_V=0$, 0.5, and 1, respectively.
Other than the varied parameter in each panel, we adopt the fiducial values
($\xi_\mathrm{crit}=10$ \AA, $n_\mathrm{c}=30$, and $\epsilon_V=0.5$).
\label{fig:coag_para}}
\end{figure}

We also examine the dependence on the parameters relevant for coagulation.
First, we present the effect of $E_\mathrm{roll}$, which regulates compaction.
As indicated in Section \ref{subsubsec:coag}, $E_\mathrm{roll}$ is determined by the critical
displacement $\xi_\mathrm{crit}$ in our model. In Fig.\ \ref{fig:coag_para}a, we show the results for
$\xi_\mathrm{crit}=2$, 10, and 30 \AA\ (note that $E_\mathrm{roll}$ is proportional to
$\xi_\mathrm{crit}$). We only show the results at $t=100$ Myr since the effect of
$\xi_\mathrm{crit}$ is
qualitatively similar at all ages.
We observe that the filling factor at $a\sim 0.1~\micron$ is affected by
the change of $\xi_\mathrm{crit}$ (or $E_\mathrm{roll}$), with larger $\xi_\mathrm{crit}$
showing smaller $\phi_m$. This is because compaction is less efficient in the case of
larger $\xi_\mathrm{crit}$. In contrast, the filling factor at $a\lesssim 0.01~\micron$ is not
affected by $\xi_\mathrm{crit}$ because compaction is not important in that grain radius range.
The filling factor also converges to the same value determined roughly by
$1/(1+\epsilon_V)$ at the largest grain radii as mentioned above.
The grain size distribution, on the other hand, is insensitive to $\xi_\mathrm{crit}$.
Since the grain abundance is dominated by large grains,
small grains collide predominantly with large grains. In this situation, the collisional
cross-section is governed by large grains which are almost compact. Thus, the
difference in the porosity at submicron radii does not affect the grain size distribution.

Grain compaction is also affected by the number of contacts $n_\mathrm{c}$
(equation \ref{eq:v120}). In Fig.\ \ref{fig:coag_para}b, we show the results for
$n_\mathrm{c}=10$, 30, and 100 at $t=100$ Myr. Naturally, the effect of $n_\mathrm{c}$
appears at grain radii where compaction is important.
As $n_\mathrm{c}$ becomes larger, the filling factor is kept lower against compaction,
so that the increase of $\phi_m$ becomes shallower at large $a_m$.
The grain size distribution is insensitive also to the change of $n_\mathrm{c}$.

We also examine the dependence on $\epsilon_V$, which regulates the maximum compaction.
As we argued above, the filling factor roughly converges to $1/(1+\epsilon_V)$ at large
grain radii, where grain velocities are high enough for significant compaction. As shown above,
the minimum value of $\phi_m$ is around 0.3--0.5. Thus, $\epsilon_V$ should be smaller than
$\sim 1$; otherwise, the filling factor at large grain radii decreases below the minimum value,
which means that we add artificial (or contradictory) porosity to the grains in compaction.
Thus, we examine $\epsilon_V=0$, 0.5, and 1 at $t=100$ Myr in Fig.\ \ref{fig:coag_para}c.
We confirm that the effect of $\epsilon_V$ appears at large
grain radii ($a\gtrsim 0.2~\micron$).
As mentioned above, $\phi_m$ at large grain radii is roughly $1/(1+\epsilon_V)$. 
The grain size distribution is affected by $\epsilon_V$ since larger porosity makes the
collision kernel larger by a factor of $\phi_m^{-1/3}$ (Section \ref{subsec:kernel}), which increases the
grain--grain collision rate.

\subsection{Pure shattering}\label{subsec:shat}

\begin{figure}
\includegraphics[width=0.48\textwidth]{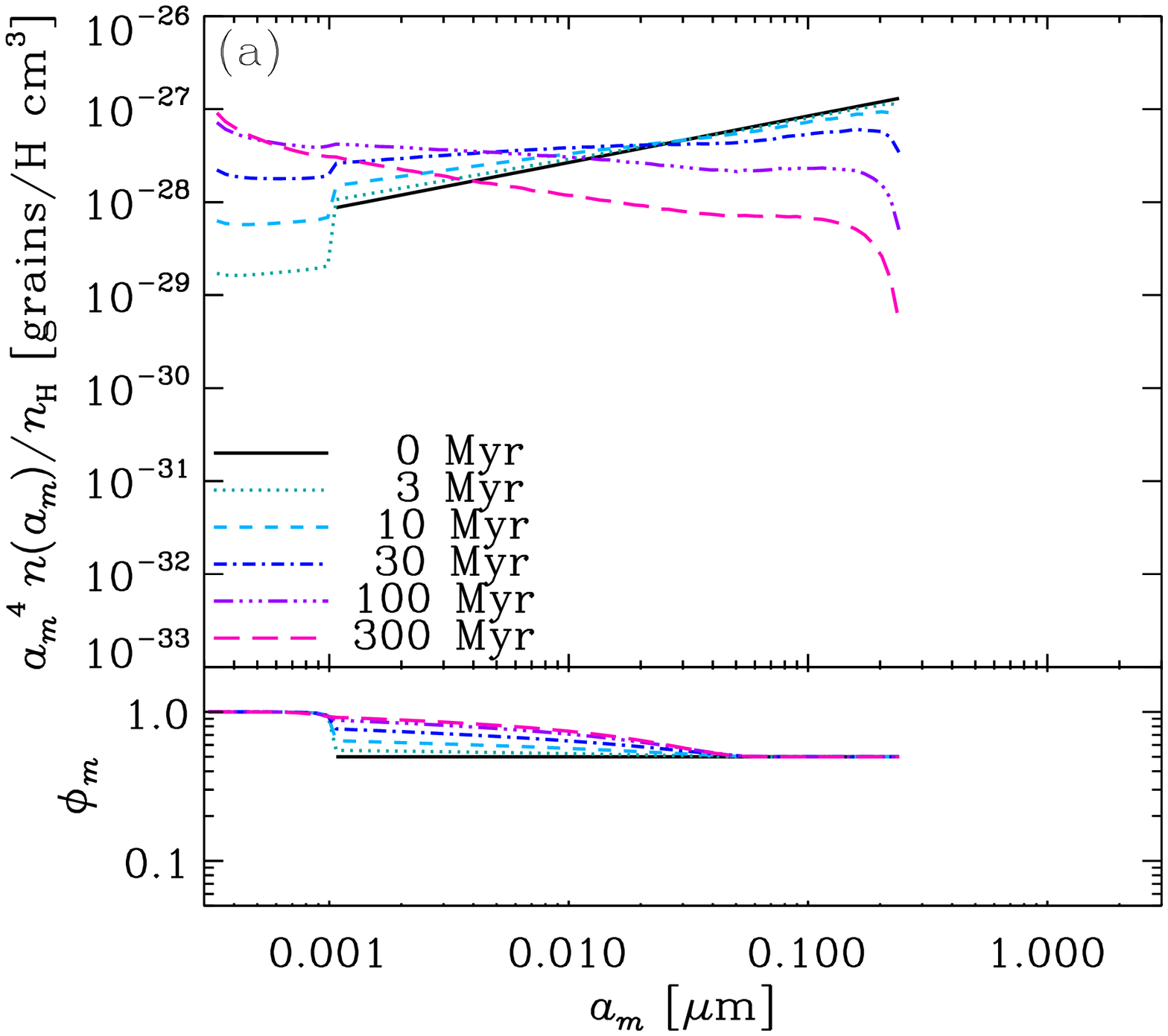}
\includegraphics[width=0.48\textwidth]{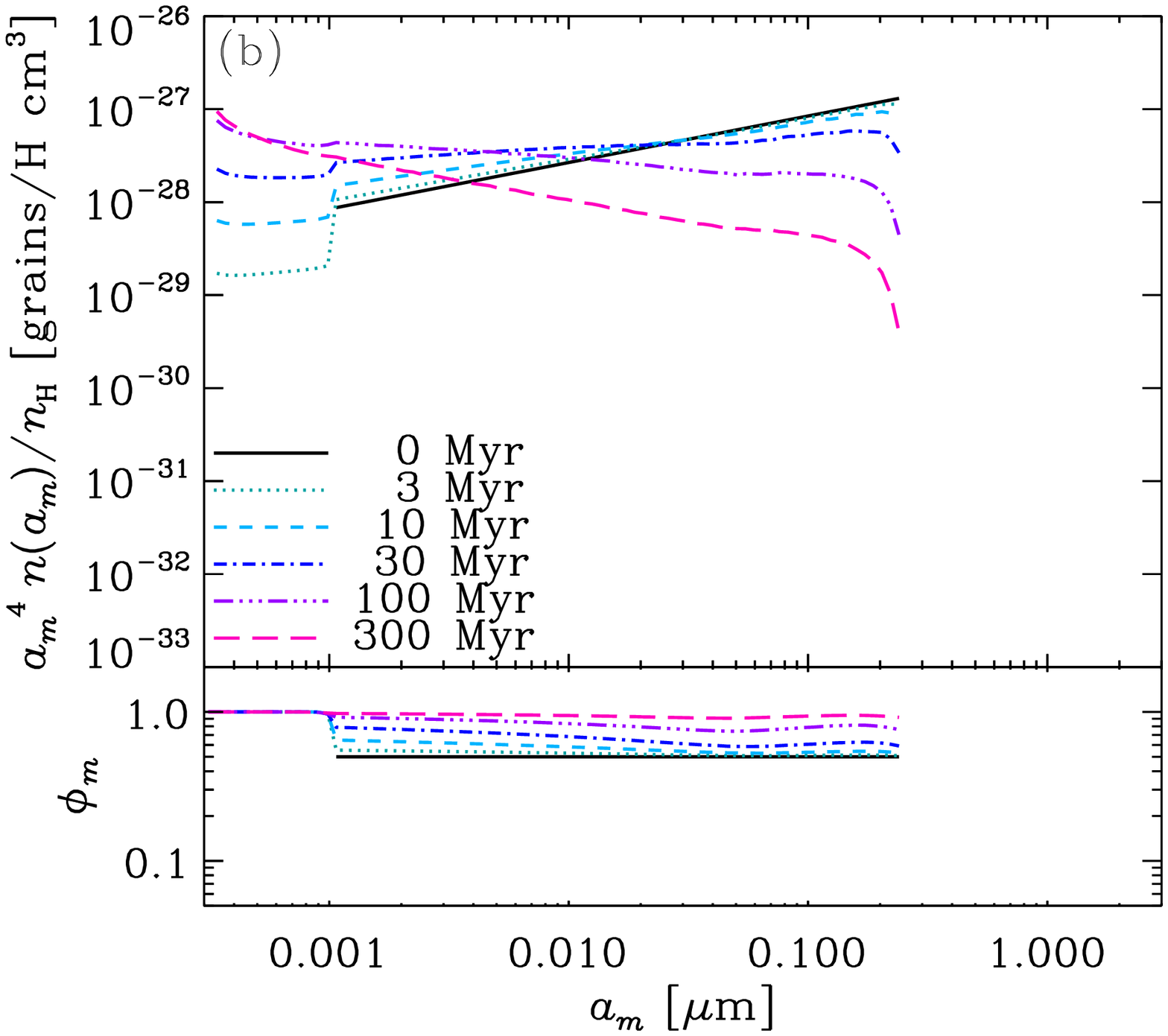}
\caption{Same as Fig.\ \ref{fig:coag} but for the pure shattering model.
We show the cases where we (a) do not consider and (b) consider compaction for the shattered remnants.
\label{fig:shat}}
\end{figure}

We show the evolution of grain size distribution and filling factor by shattering in the diffuse ISM.
Since shattering does not create new porosity in grains, we start with $\phi_m =0.5$ for all grains.
(If we choose $\phi_m=1$ as the initial condition, the filling factor is always 1.)
We show the results
in Fig.\ \ref{fig:shat}a. Shattering continuously converts large grains to small grains, creating
grains down to the minimum grain radius $a=3$ \AA.
After $t\sim 100$ Myr, the grain abundance is dominated by small grains.
The filling factor increases, especially at small grain radii, because shattered fragments are
compact. Since we assume that the remnants have the same filling factor as the original
grains, the filling factors of the largest grains, which are dominated by shattered remnants,
do not change. Recall that the largest fragment size is $(0.02)^{1/3}\simeq 0.27$ times
the original grain size (Section \ref{subsubsec:shat}).
Since the maximum grain radius in the initial condition is 0.25 $\micron$,
the largest grain radius where $\phi_m$ is raised by fragments is
$a\simeq 0.27\times 0.25~\micron\simeq 0.068~\micron$. Thus, the porosity is only changed
from its initial value at $a<0.068~\micron$. All the grains with $a<a_\mathrm{min,ini}=0.001~\micron$
are newly formed by shattering; thus, they always have $\phi_m=1$.

As explained in Section \ref{subsubsec:param_shat}, the shattered remnants could suffer compaction.
In order to examine this effect, we also show the case with compaction of remnants;
that is, we adopt equation~(\ref{eq:compact_rem}) instead of the first condition in
equation~(\ref{eq:V12_shat}) for the remnant volume.
In Fig.\ \ref{fig:shat}b, we show the evolution of grain size distribution and filling factor
with including remnant compaction. The grain size distributions are little affected
by the treatment of the remnant volume, while the filling factor at large grain radii
increases if we take compaction of remnants into account.
Because shattering never creates porosity, the filling factors of all grains converge
to unity on a time-scale of depleting large grains by shattering (100--300 Myr).

\subsection{Coagulation and shattering}\label{subsec:coag_shat}

\begin{figure}
\includegraphics[width=0.48\textwidth]{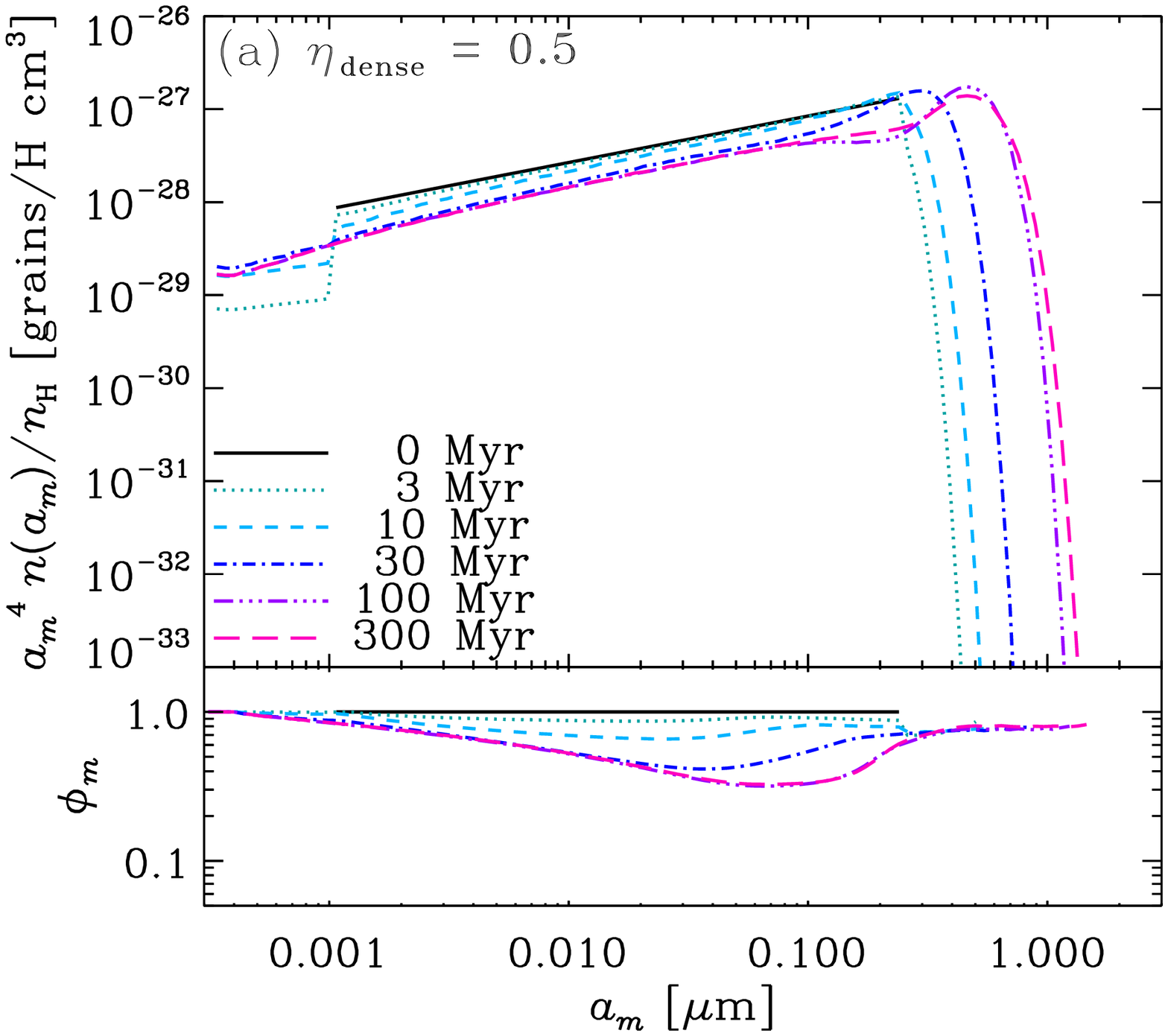}
\includegraphics[width=0.48\textwidth]{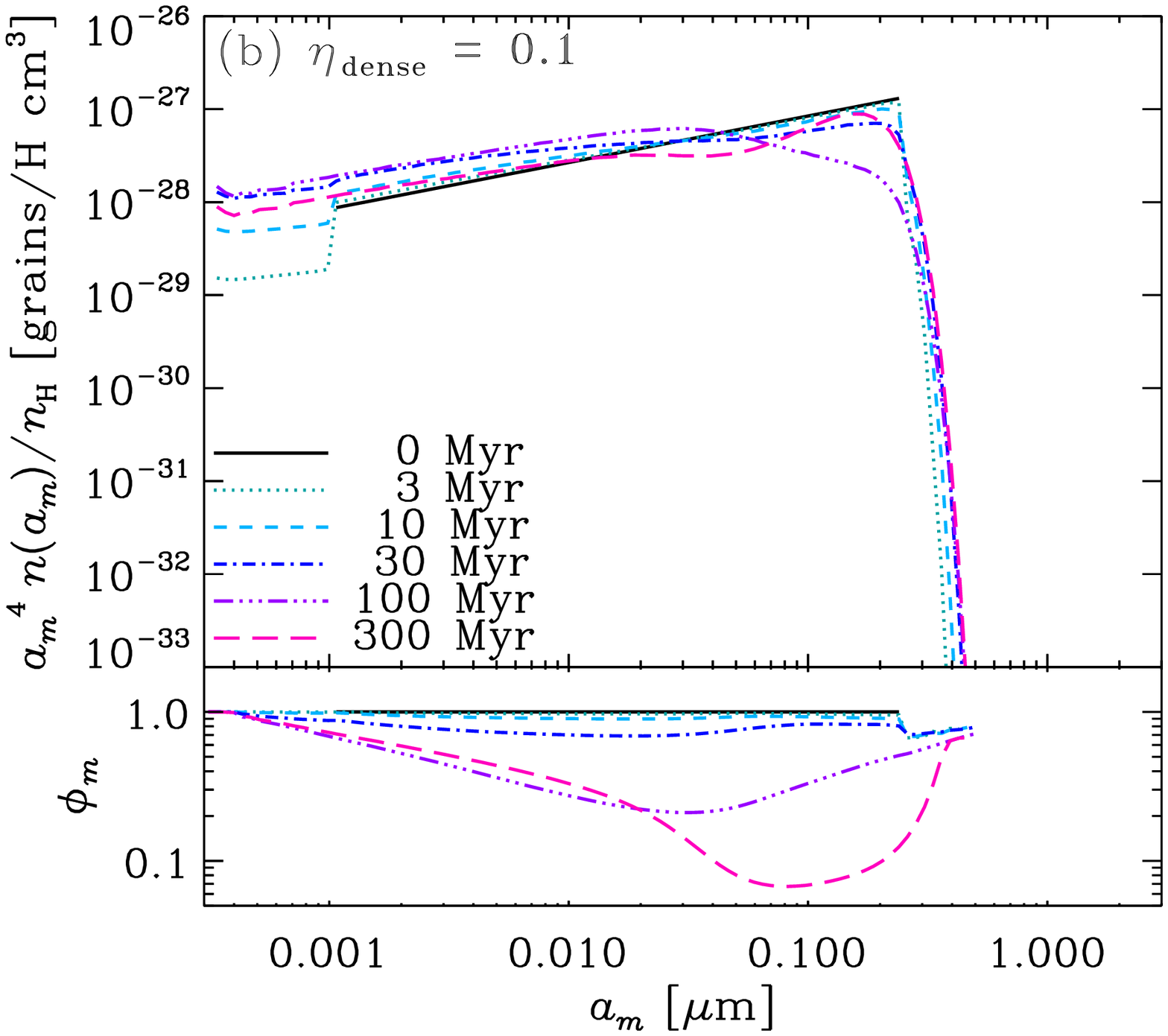}
\includegraphics[width=0.48\textwidth]{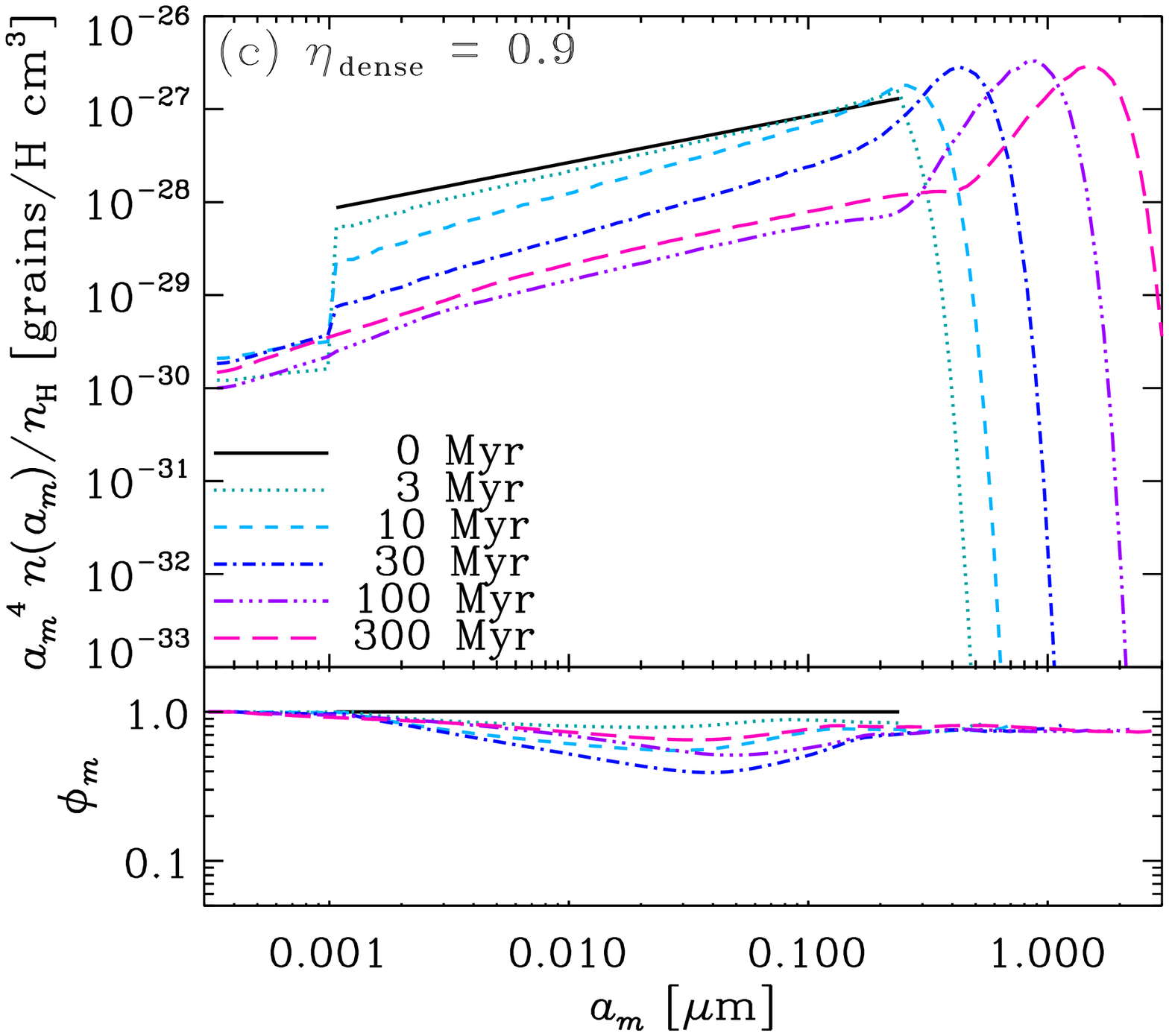}
\caption{Same as in Fig.\ \ref{fig:coag}, but including both shattering and coagulation.
Panels (a), (b) and (c) show the results for $\eta_\mathrm{dense}=0.5$ (fiducial),
0.1, and 0.9, respectively. Note the different scale of $\phi_m$ between this figure and
Fig.\ \ref{fig:coag}.
\label{fig:coag_shat}}
\end{figure}

If we consider the grain size distribution in a region large enough to include both the dense
and diffuse ISM (or if we consider time-scales much longer than the phase-exchange time;
Section~\ref{subsec:param}),
the effect of coagulation and that of shattering coexist.
In our one-zone model, we could simulate this coexisting effect by simultaneously treating
coagulation and shattering at each time-step with a weight of
$\eta_\mathrm{dense} : (1-\eta_\mathrm{dense})$
as formulated in equations (\ref{eq:rho_mix}) and (\ref{eq:psi_mix}).
We adopt $\eta_\mathrm{dense}=0.5$ first as a fiducial value (that is, the grains spend half of
their times in the dense ISM).
In Fig.~\ref{fig:coag_shat}a, we show the evolution of grain size distribution and filling factor.

We observe in Fig.~\ref{fig:coag_shat}a that, if both coagulation and shattering are present,
the grain size distribution maintains a power-law-like shape.
It is also interesting to note that the slope is similar to that of the MRN distribution.
It has been shown that the slope of grain size distribution converges to a value similar to
the MRN distribution if efficient fragmentation and/or coagulation
occur \citep[e.g.][]{Dohnanyi:1969aa,Williams:1994aa,Tanaka:1996aa,Kobayashi:2010aa}.
Coagulation is slightly `stronger' than shattering, so that large (submicron) grains gradually increase.
The grain size distribution and filling factor reach an equilibrium
around $t\sim 100$~Myr. The filling factor changes in a way similar to the pure coagulation
case (Fig.\ \ref{fig:coag}) but the decrease of the filling factor proceeds more if both
coagulation and shattering are present because small grains which coagulate to
form porosity are continuously supplied. The filling factor at large grain radii converges
to $\sim 1/(1+\epsilon_V)$ as explained in Section \ref{subsec:coag}.

To examine the balance between coagulation and shattering, we also show
the results for different values of $\eta_\mathrm{dense}$ (0.1 and 0.9) in
Fig.~\ref{fig:coag_shat}.
Naturally, a higher abundance of large grains is obtained for larger $\eta_\mathrm{dense}$.
For $\eta_\mathrm{dense}=0.9$, the result is similar to
that in the pure coagulation case (Fig.~\ref{fig:coag}), except that small grains continue to be produced
by shattering.
The case of $\eta_\mathrm{dense}=0.1$ has the following two interesting properties: (i)
The evolution of grain size distribution is not monotonic. Indeed, large grains
are continuously converted into small grains up to $t\sim 100$~Myr but
large grains increase (`re-form') after that. (ii) The filling factor becomes very small at
$t>100$~Myr.
This is because the production of small grains by shattering continuously activates coagulation,
which creates porosity. The increased porosity makes the geometrical cross-section larger, and
enhances coagulation. Moreover, since the velocities of porous grains are reduced,
compaction does not occur as efficiently as in the pure coagulation case.
Thus, efficient shattering together with coagulation makes the interstellar dust
highly porous and further activates coagulation.
This interesting interplay is further discussed in Section \ref{sec:discussion}.

\begin{figure}
\includegraphics[width=0.48\textwidth]{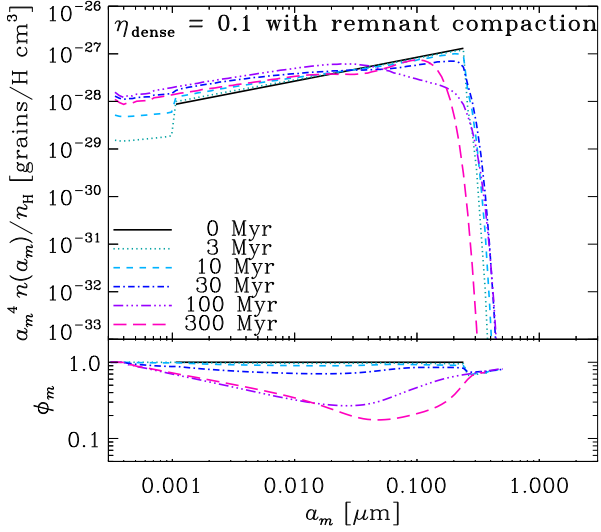}
\caption{Same as Fig.\ \ref{fig:coag_shat}b but with compaction of shattered remnants.
\label{fig:coag_shat_compact}}
\end{figure}

The dependences on the parameters related to coagulation
($\xi_\mathrm{crit}$, $n_\mathrm{c}$ and $\epsilon_V$) are similar to those shown
in Section~\ref{subsec:coag} (Fig.~\ref{fig:coag_para}).
On the other hand, compaction of shattered remnants (Section~\ref{subsubsec:param_shat})
could decrease the porosity if $\eta_\mathrm{dense}$ is low.
In Fig.\ \ref{fig:coag_shat_compact}, we show the results with
compaction of shattered remnants for $\eta_\mathrm{dense}=0.1$.
This figure should be compared with Fig.\ \ref{fig:coag_shat}b.
We observe that the filling factor becomes higher at $t\sim 300$ Myr at large grain radii,
although the results are almost unchanged at younger ages. Compaction of remnants
also makes coagulation at later epochs less efficient, producing less large grains at
$t\sim 300$ Myr. However, the filling factor is still as small as 0.1--0.2 at $a\sim 0.1~\micron$
at $t\sim 300$ Myr; thus, the conclusion that efficient shattering together with coagulation
helps to increase porosity is robust.

\section{Extinction curves}\label{sec:ext_result}

Based on the grain size distributions and the filling factors presented above,
we calculate the extinction curves by the method in Section~\ref{subsec:ext_method}
for silicate and amC. As mentioned above,
the extinction $A_\lambda$ is presented in two ways: $A_\lambda /N_\mathrm{H}$ and
$A_\lambda /A_V$. To clarify the effect of porosity, we also calculate extinction curve by
forcing the filling factor of all grains to be unity with $a_m$ fixed.
The extinction calculated in this way (i.e.\
$\phi_m =1$) is denoted as $A_{\lambda ,1}$. The effect of porosity is presented by
$A_\lambda /A_{\lambda ,1}$.

\subsection{Effect of coagulation}

\begin{figure}
\begin{center}
\includegraphics[width=0.45\textwidth]{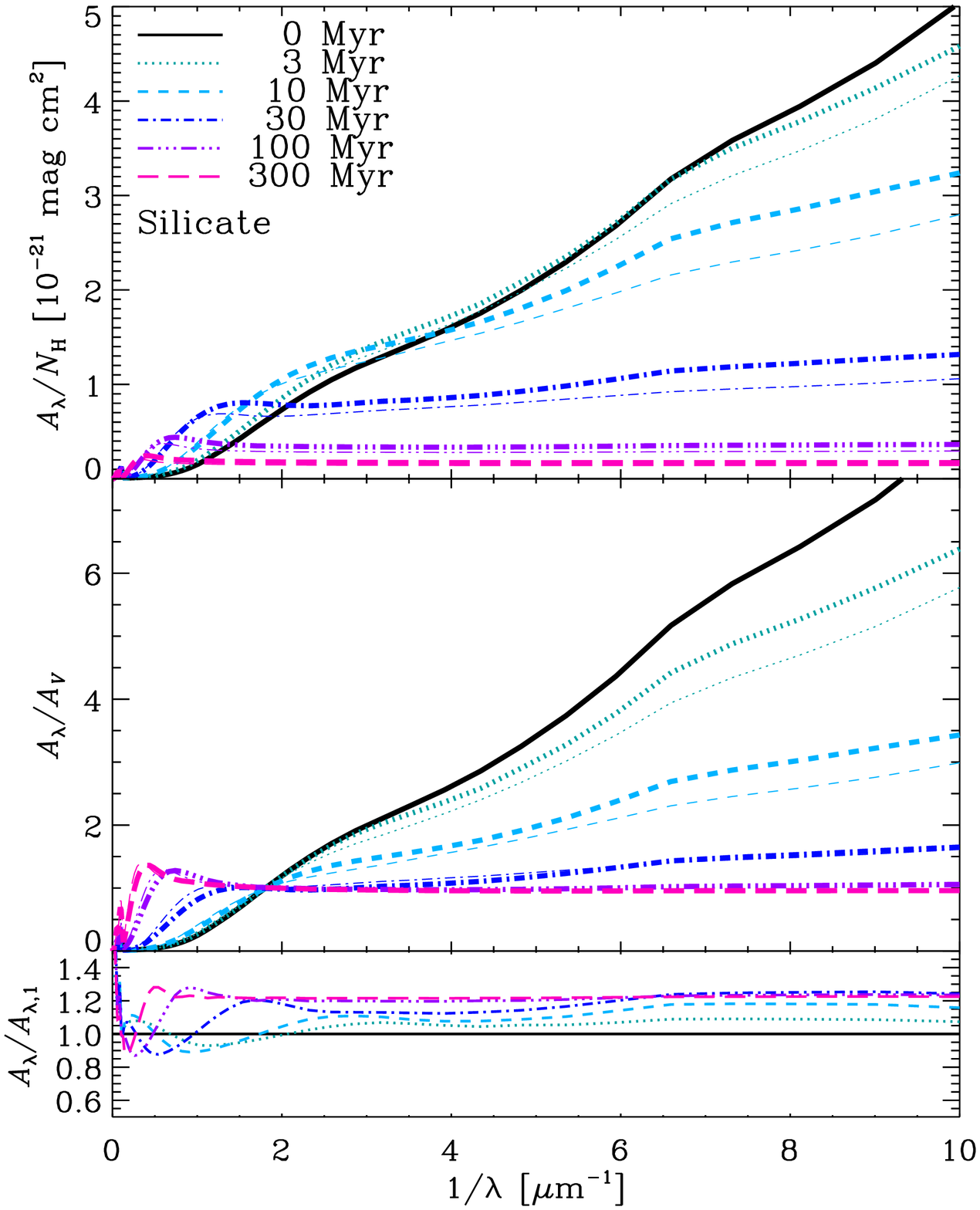}
\includegraphics[width=0.45\textwidth]{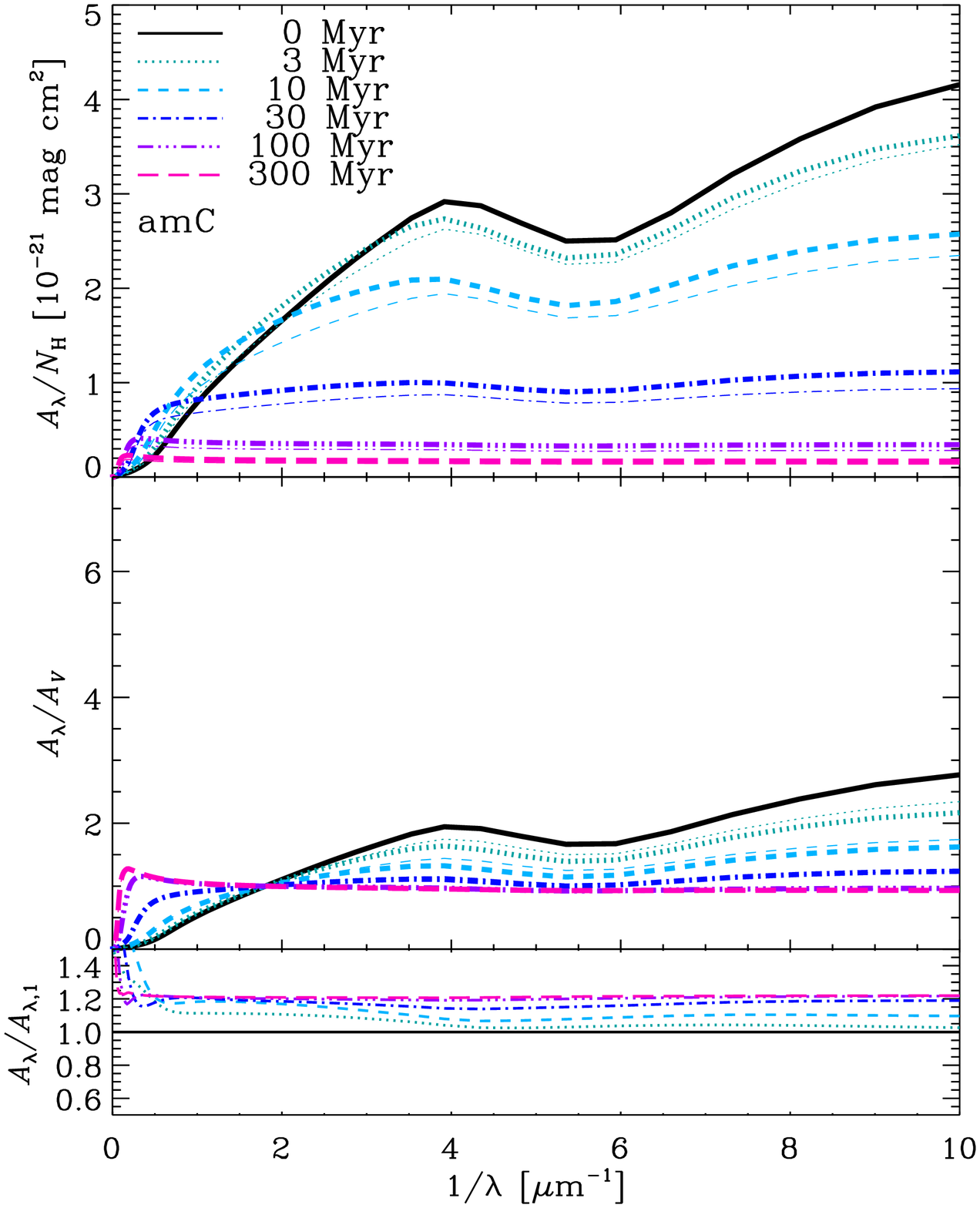}
\end{center}
\caption{Extinction curves for silicate and amC in the upper and lower panels, respectively.
The solid, dotted, short-dashed, dot--dashed, triple-dot--dashed, and long-dashed lines (thick lines)
show the extinction curves at $t=0$ (initial condition), 3, 10, 30, 100, and 300 Myr, respectively.
The thin lines with the same line species show $A_{\lambda ,1}$ (extinction with $\phi_m=1$).
Note that
$A_\lambda =A_{\lambda ,1}$ at $t=0$.
In each panel, the upper, middle, and lower windows present the extinction per hydrogen,
the extinction normalized to the $V$-band value, and the ratio of $A_\lambda$ to
$A_{\lambda ,1}$ (an indicator of the porosity effect), respectively.
\label{fig:ext_coag}}
\end{figure}

In Fig.~\ref{fig:ext_coag}, we present the extinction curves corresponding to the
grain size distributions and the filling factors for the pure coagulation case (Fig.~\ref{fig:coag}).
As expected from the evolution of grain size distribution, the extinction curve becomes flatter
as coagulation proceeds. Since a large fraction of the grains become bigger than
$\sim 0.1~\micron$ after coagulation,
the extinction per hydrogen declines at wavelengths shorter than $\sim 2\upi\times 0.1~\micron$
for both materials. The effect of porosity, which appears in the difference between
$A_\lambda$ (thick lines) and $A_{\lambda ,1}$ (thin lines) or
in the ratio $A_\lambda /A_{\lambda ,1}$ shown in the bottom window, is clear.
The extinction is 10--20 per cent higher in the
porous case than in the non-porous case in a large part of wavelengths for both materials.

There are some differences between the two materials.
For silicate, porous grains have smaller extinction than compact ones in a certain wavelength range
(e.g.\ $1/\lambda\sim 0.5$--2~$\micron^{-1}$ at $t=10$~Myr) and this range shifts to
longer wavelengths with age.
\citet{Voshchinnikov:2006aa} and \citet{Shen:2008aa} showed that
the porosity could both increase and decrease the
extinction cross-section depending on the wavelength. The wavelength range of suppressed
extinction by porosity
is roughly consistent with their results. \citet{Voshchinnikov:2006aa} also showed that
the wavelength range where the extinction cross-section
is decreased by porosity shifts to longer wavelengths as the grains
become larger. This is consistent with our result.
At $t\lesssim 100$ Myr, the normalized extinction curves ($A_\lambda /A_V$) becomes steeper by the
porosity because the increase of extinction is more enhanced in the UV than in the $V$ band.

For amC, porosity always increases the extinction; that is,
$A_\lambda /A_{\lambda ,1}$ is always larger than 1. Thus,  if we normalize the extinction to $A_V$,
the porosity effect roughly cancels out, so that the extinction curve shape is insensitive to
the porosity (filling factor) for amC.


We also examined the parameter dependence (not shown) and confirmed that
the extinction curves are
not sensitive to $\xi_\mathrm{crit}$ or $\epsilon_V$ in the optical and UV with differences
less than 5 per cent.
However, at $1/\lambda <2~\micron^{-1}$,
the difference could be as large as 20 per cent, since, at such long wavelengths,
the porosity of large grains, which is
regulated by $\xi_\mathrm{crit}$ and $\epsilon_V$, is important.
The changes driven by the difference in those parameters
are sub-dominant compared with the difference between porous and non-porous grains
shown in Fig.\ \ref{fig:ext_coag}.

\subsection{Inclusion of shattering}

In the pure shattering case, the filling factor simply tends to increase; thus, the resulting
extinction curve approaches the one with compact grains. More important is the interplay between
shattering and coagulation as shown above. Indeed, coagulation of
newly created small grains by shattering produces porous grains, contributing to the
increase of porosity in the interstellar dust. Here, we investigate the effect of relative
strength between shattering and coagulation; that is, we compare the results for different
dense gas fractions, $\eta_\mathrm{dense}$.

\begin{figure}
\begin{center}
\includegraphics[width=0.45\textwidth]{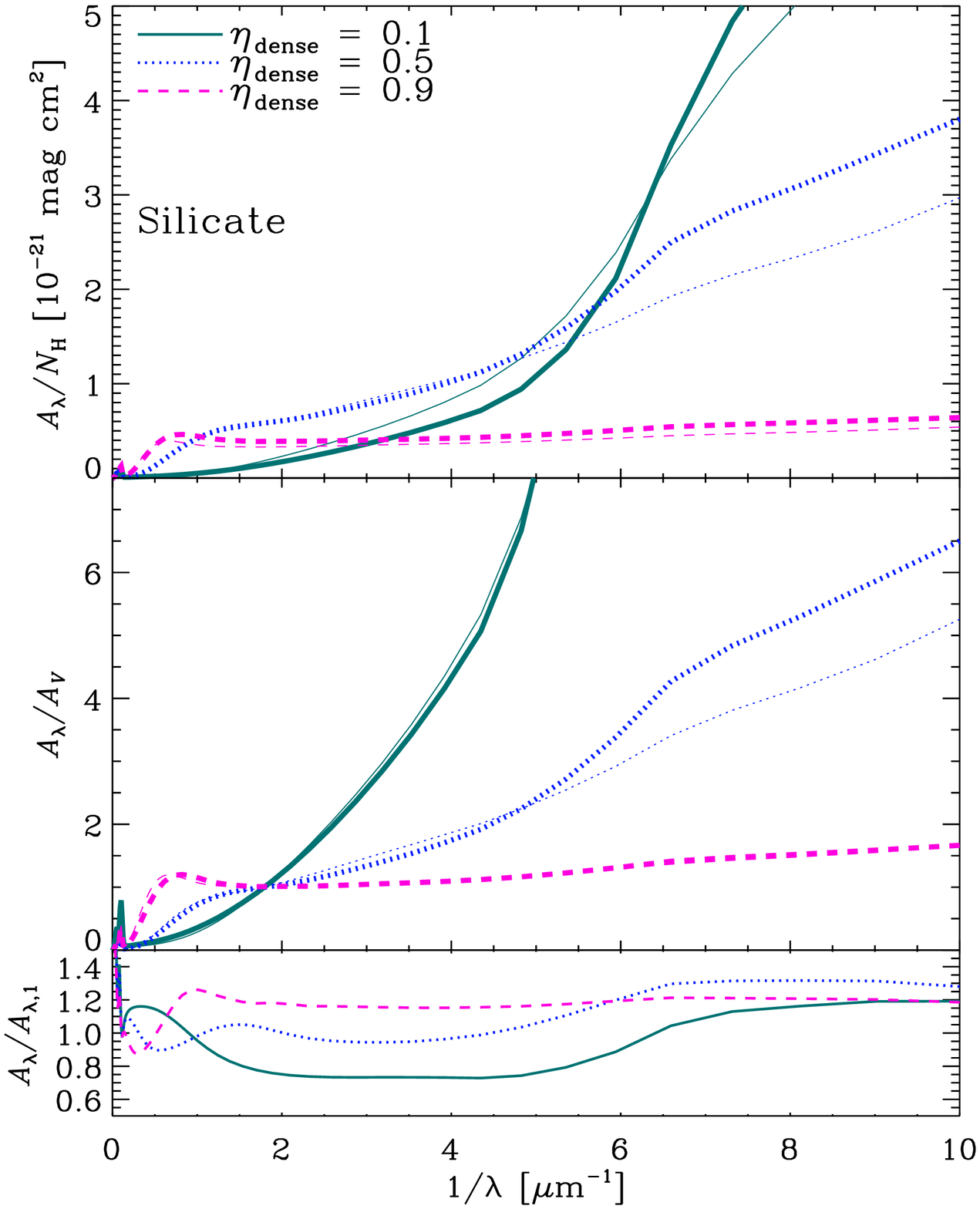}
\includegraphics[width=0.45\textwidth]{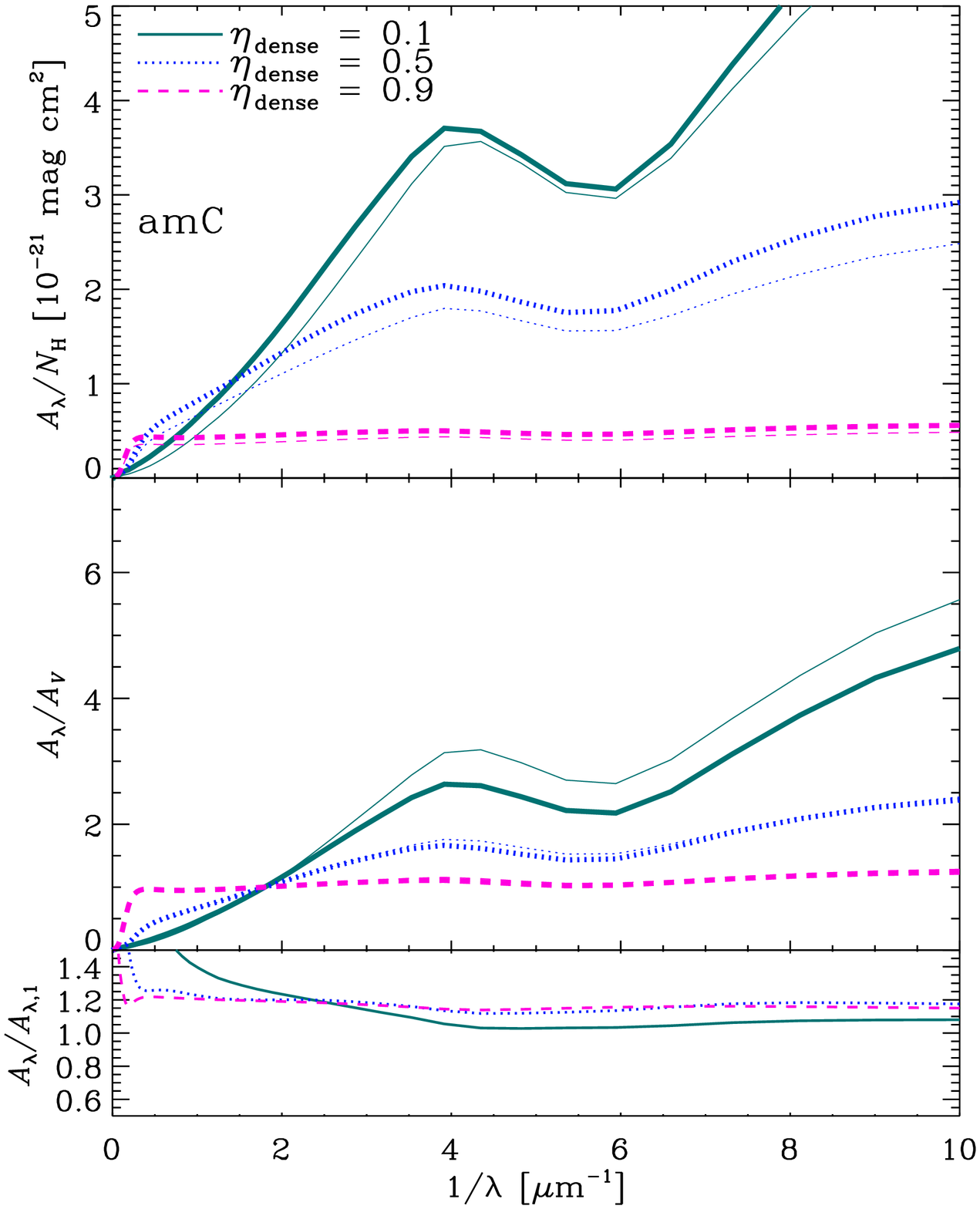}
\end{center}
\caption{Comparison among the extinction curves in the models including
both coagulation and shattering. The relative strengths of these two processes are regulated by
the dense gas fraction $\eta_\mathrm{dense}$. We show the same quantities as in Fig.\ \ref{fig:ext_coag}
at $t=100$ Myr.
The solid, dotted, and dashed lines show the results for $\eta_\mathrm{dense}=0.1$, 0.5, and 0.9,
respectively. The thick and thin lines present the cases using the calculated filling factors
(Fig.\ \ref{fig:coag_shat}) and those with $\phi_m=1$ (the filling factor is forced to be unity; i.e.\ $A_{\lambda ,1}$).
The upper and lower panels are for silicate and amC, respectively.
\label{fig:coag_shat_comp}}
\end{figure}

In Fig.\ \ref{fig:coag_shat_comp}, we compare the extinction curves calculated for
the models including both coagulation
and shattering. The corresponding grain size distributions are shown in
Section \ref{subsec:coag_shat} (Fig.\ \ref{fig:coag_shat}).
We compare the results at the same age $t=100$ Myr. We observe that
the steepness of extinction curve is very sensitive to $\eta_\mathrm{dense}$.
This is a natural consequence of the different grain size distributions for various
$\eta_\mathrm{dense}$.
We also plot the extinction curves with $\phi_m=1$ (with the same distribution of $a_m$),
that is, $A_{\lambda ,1}$. Comparing $A_\lambda$ with $A_{\lambda ,1}$,
we observe for silicate that the porosity increases the
extinction at short and long wavelengths, but that it rather decreases the extinction at
intermediate wavelengths as already noted in the previous subsection.
The wavelength range where the porosity decreases the extinction shifts to shorter wavelengths
for smaller $\eta_\mathrm{dense}$,
which is consistent with the above mentioned tendency that porosity of smaller grains reduces 
the extinction at shorter wavelengths. For amC, the porosity always increases the extinction,
but the largest porosity
(i.e.\ $\eta_\mathrm{dense}=0.1$) does not necessarily mean the largest opacity enhancement
($A_\lambda /A_{\lambda ,1}$) in the UV.
At long wavelengths, larger porosity indicates larger extinction.

Overall, the effect of porosity on the extinction curve is less than 20
per cent as far as the UV--optical extinction curves are concerned, but this could mean that we
`save' up to 20 per cent of metals to realize an observed extinction if we take porosity into account.
{The difference is not large, though, because the increase of grain volume and the
`dilution' of permittivity have opposite effects on the extinction \citep{Li:2005aa}.
For further constraint on the porosity evolution,}
we need a comprehensive
analysis of observed dust properties including the infrared SED
\citep{Dwek:1997aa}. Polarization may also help to further constrain the models.
Since such a detailed comparison also needs further modelling of the mixture of various
dust species and of the interstellar radiation field, we leave it for a future work.

\section{Discussion}\label{sec:discussion}

\subsection{Effects of porosity on the evolution of grain size distribution}

The porosity increases the effective sizes of grains, enhancing the grain--grain collision rate.
If only coagulation occurs, however, the porosity, which is large at $a\lesssim 0.1~\micron$, does not
have a large influence on the grain size distribution, since grains quickly coagulate
to achieve $a\gtrsim 0.1~\micron$, and are affected by compaction.
Note that, if we individually observe clouds denser than $n_\mathrm{H}\sim 10^3$ cm$^{-3}$,
we could find very fluffy grains
as will be shown by L. Pagani et al.\ (in preparation)
\citep[see also][]{Hirashita:2013aa,Wong:2016aa}. The major part of interstellar grains are
smaller than 1 $\micron$ \citep[e.g.][]{Weingartner:2001aa}, so that the grains in such dense
clouds are not likely to contribute directly to the interstellar dust population.

Coagulation can be balanced by shattering.
For $\eta_\mathrm{cold}=0.5$, the grain size distribution and the filling factor both converge to
equilibrium distributions at $t\sim 100$ Myr, while coagulation continuously increases the grain radii
for $\eta_\mathrm{cold}=0.9$
(Section \ref{subsec:coag_shat}; Fig.\ \ref{fig:coag_shat}).
Thus, cases with high $\eta_\mathrm{cold}$ produce
similar results to the pure coagulation case.
For the case of $\eta_\mathrm{cold}=0.5$, we performed a calculation by forcing $\phi_m$ to be
always unity (not shown) and found that the grain size distribution changes little by the different
treatment of $\phi_m$ (i.e.\ compared with the results shown in Fig.\ \ref{fig:coag_shat}a).
This is because the
evolution of grain size distribution is regulated by the formation of large grains
(recall that the grain abundance is dominated by large grains), which have small porosity
because of compaction.

\begin{figure}
\begin{center}
\includegraphics[width=0.45\textwidth]{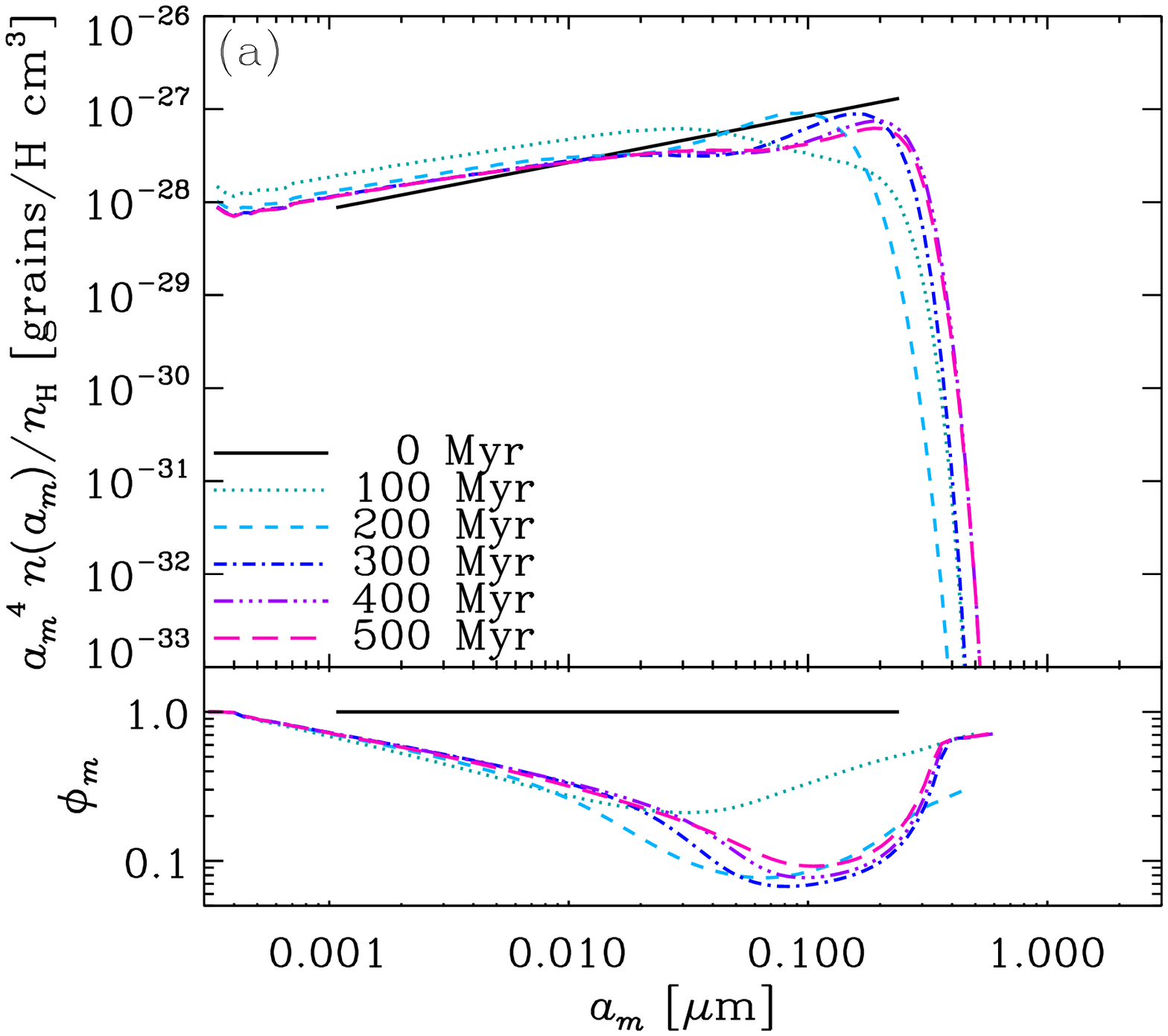}
\includegraphics[width=0.45\textwidth]{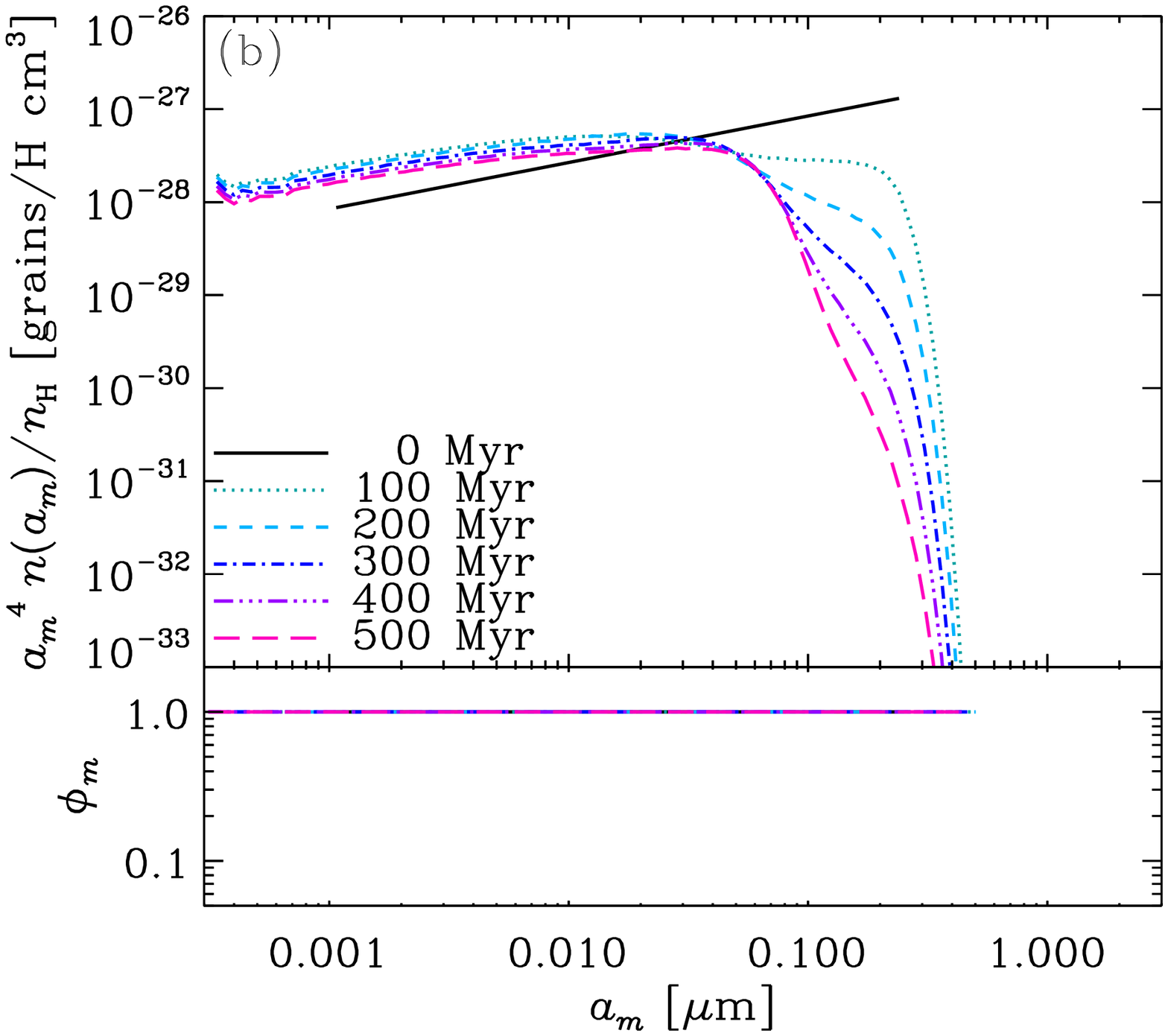}
\end{center}
\caption{Long-term evolution of grain size distribution and filling factor
with both coagulation and shattering under $\eta_\mathrm{dense}=0.1$.
Panels (a) and (b) show the results with including the evolution of filling factor
(i.e.\ the same model as in Fig.\ \ref{fig:coag_shat}b) and with fixing $\phi_m=1$,
respectively. We present the
results at $t=0$, 100, 200, 300, 400, and 500 Myr (at $t<100$ Myr, the two results have
little difference) by the
solid, dotted, short-dashed, dot--dashed, triple-dot--dashed, and long-dashed lines, respectively.
\label{fig:coag_shat_eta01}}
\end{figure}

Porosity plays a critical role in the case of small $\eta_\mathrm{dense}=0.1$.
As discussed in Section~\ref{subsec:coag_shat} (Fig.~\ref{fig:coag_shat_compact}),
the grain size distribution is first dominated by shattering, which decreases the abundance
of large grains, while coagulation gradually recovers the large-grain abundance at later
stages (typically after $t\sim 100$~Myr). We argued above that coagulation is
activated at $t\sim 100$~Myr because the increased porosity enhances the grain cross-sections
(the grain--grain collision rate).
For demonstration, we compare two calculations in Fig.\ \ref{fig:coag_shat_eta01}:
one is the same as above for
$\eta_\mathrm{dense}=0.1$ (Fig.\ \ref{fig:coag_shat}b), and the other is a calculation
with the same setting but with
$\phi_m=1$ (i.e.\ without porosity). Since the difference is not prominent on a short time-scale,
we focus on the evolution after $t=100$ Myr. If we fix $\phi_m=1$, grains at
$a\gtrsim 0.1~\micron$ are simply depleted by shattering as expected from
weak coagulation in $\eta_\mathrm{dense}=0.1$. In contrast, if we take the evolution of $\phi_m$
into account, grains at $a\gtrsim 0.1~\micron$
are re-formed at $t\gtrsim 200$ Myr, reaching roughly an equilibrium
at $t\sim 400$ Myr. The filling factor reaches the smallest value ($\phi_m\sim 0.1$) at
$a\sim 0.1~\micron$. There are two effects that promote coagulation at later stages: (i)
The increase of grain cross-sections by porosity enhances the grain--grain collision rate
because the collision kernel scales with the porosity as
$\propto\phi_m^{-1/3}$ (Section \ref{subsec:kernel}). (ii) The decreased grain velocity
($\propto\phi_m^{1/3}$; equation \ref{eq:vel}) makes compaction in coagulation at large grain
radii less effective, so that the filling factor is kept small even at $a\sim 0.1~\micron$.
These two effects are persistent qualitatively
even if we consider the compaction of shattered remnants (see Fig.\ \ref{fig:coag_shat_compact}).

Recall that our treatment of the two-phase ISM is based on the parameter $\eta_\mathrm{dense}$,
which sets the fraction of time dust spends in the dense phase.
Thus, the above results imply that the dust evolution in a condition where both coagulation
and shattering coexist (in a wide area of the ISM and/or on a time-scale longer than the phase
exchange time-scale $\sim 10^7$~yr; Section \ref{subsec:param}) could be strongly
affected by porosity. In other words, dust evolution models which do not include
porosity evolution could predict a very different evolution of the grain size distribution from
those which correctly take the porosity evolution into account.

For a long-term evolution of dust in the ISM, dust enrichment by stellar ejecta,
dust growth by the accretion of gas-phase metals, and dust destruction by supernova shocks
are also important \citep[e.g.][]{Dwek:1998aa}. Thus, it is interesting to additionally model
the porosity evolution in these processes. This paper provides a first
important step for the understanding of porosity evolution since
coagulation and shattering are mechanisms of
efficiently modifying the grain size distribution. Indeed, \citet{Hirashita:2019aa}
emphasized the importance of these two processes in realizing MRN-like
grain size distributions \citep[see also][]{Aoyama:2020aa}.

\subsection{Effects of porosity on extinction curves}\label{subsec:ext_discussion}

As shown in Section \ref{sec:ext_result}, porosity affects the UV--optical extinction curves
by 10--20 per cent. As noted by \citet{Voshchinnikov:2006aa}, porosity does not necessarily
increase the extinction \citep[see also][]{Jones:1988aa,Li:2005aa}.
At long-optical and near infrared wavelengths, the opacity of silicate can decrease
owing to the porosity.
The wavelength range where this decrease occurs shifts towards shorter wavelengths as the typical
grain radius becomes smaller in e.g.\ a shattering-dominated condition.
The extinction of amC is enhanced by 10--20 per cent at all wavelengths.
The above results indicate that we could save
10--20 per cent of dust to explain the opacity, especially at long (far-infrared) and short (UV) wavelengths.

\begin{figure}
\begin{center}
\includegraphics[width=0.45\textwidth]{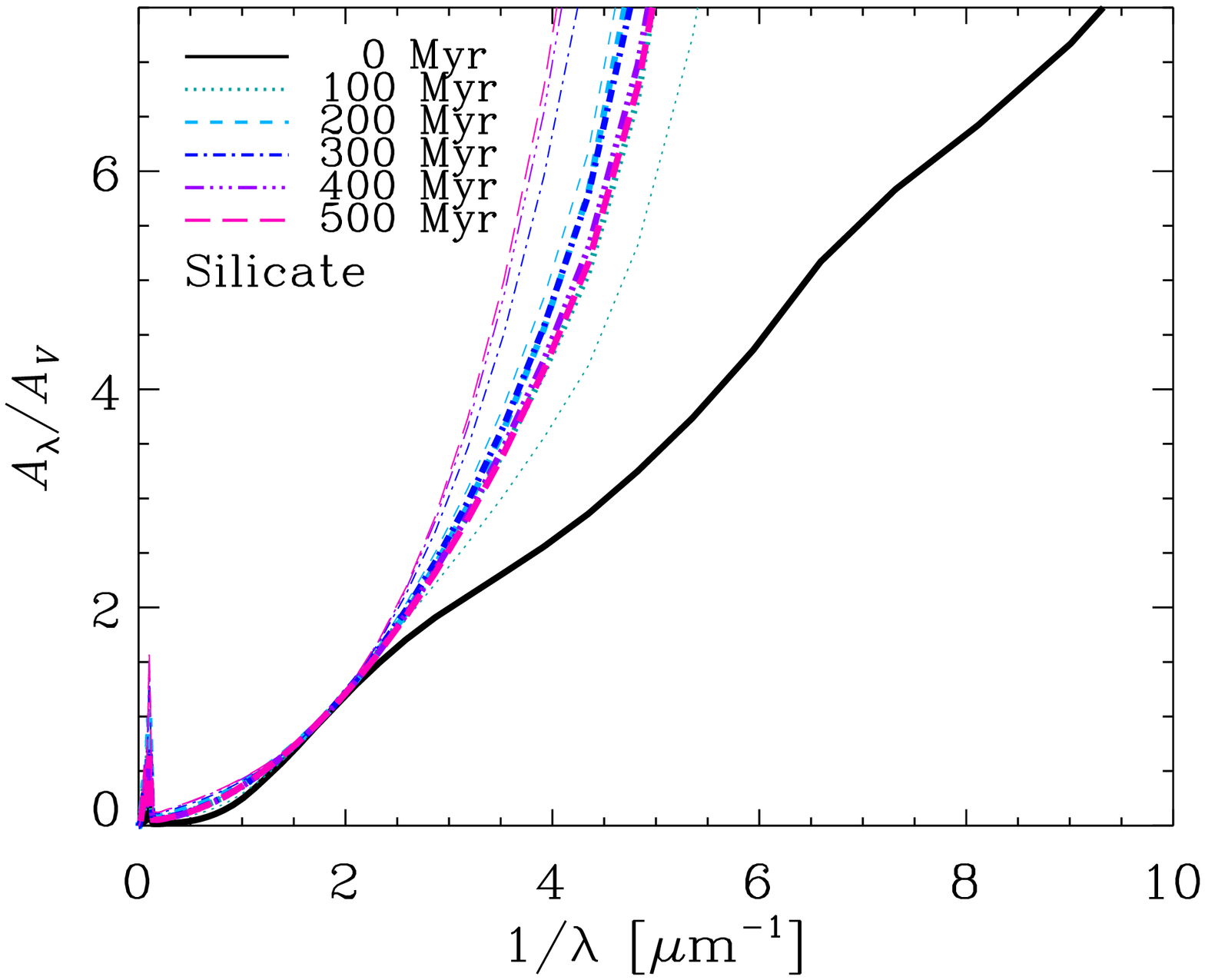}
\includegraphics[width=0.45\textwidth]{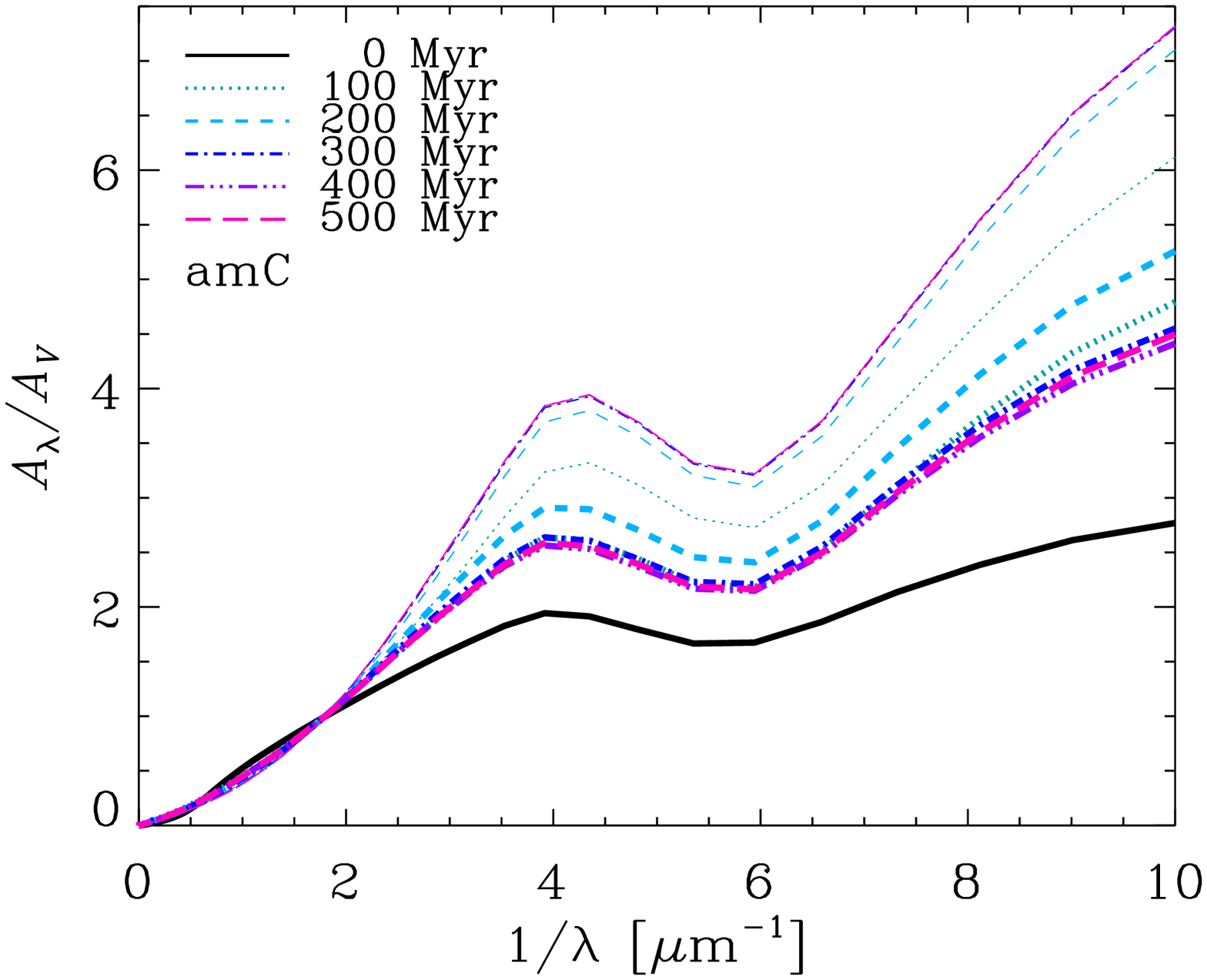}
\end{center}
\caption{Evolution of extinction curve for the model with $\eta_\mathrm{dense}=0.1$
shown in Fig.~\ref{fig:coag_shat_eta01}. Upper and lower panels
present the results for silicate and amC, respectively. The correspondence between the
line species and the age is the same as in Fig.\ \ref{fig:coag_shat_eta01}.
The thick and thin lines show the results with the evolution of porosity
(Fig.\ \ref{fig:coag_shat_eta01}a) and with forcing $\phi_m$ to be always unity
(Fig.\ \ref{fig:coag_shat_eta01}b). Note that $\phi_m=1$ at $t=0$ for both cases.
\label{fig:ext_coag_shat_eta01}}
\end{figure}

The effects of porosity on extinction curves are further pronounced given that
the evolution of grain size distribution is affected by the filling factor.
As shown above, porosity has the largest effect on the grain size distribution if
strong shattering and weak coagulation are both present such as in the case of
$\eta_\mathrm{dense}=0.1$. Thus, the extinction curve is affected by porosity not only through the
optical properties but also through the modification of grain size distribution.
To present this effect, we show in Fig.\ \ref{fig:ext_coag_shat_eta01}
the extinction curves corresponding to the cases shown in Fig.\ \ref{fig:coag_shat_eta01}.
On such a long time-scale as shown in Fig.\ \ref{fig:coag_shat_eta01},
the loss of dust through the lower grain radius boundary by shattering
(recall that we remove the grains which become smaller than $a=3$~\AA)
leads to a decrease of $A_\lambda /N_\mathrm{H}$ just because of the boundary condition.
Thus, we only show $A_\lambda /A_V$, which is free from the effect of dust mass loss
(i.e.\ we could purely observe the effect of grain size distribution).
For the case without porosity evolution ($\phi_m=1$), the extinction curve continues to
be steepened because shattering continues to
convert large grains to small grains. If the porosity evolution is included,
the extinction curve is flattened with age because large
grains are `recreated' by enhanced coagulation at later times.
Both species show steeper extinction curves than
the initial state for the case with porosity evolution although the grain size distributions
are similar to the initial condition (Fig.\ \ref{fig:coag_shat_eta01}).
This is because the grain opacity does not increase in the $V$ band, where the
extinction curve is normalized, while
porosity enhances the UV extinction by 30--40 per cent.
\citet{Jones:1988aa} and \citet{Voshchinnikov:2006aa} also showed that porosity does not
change the optical extinction
much but increase the UV and infrared opacities.
This effect makes the UV extinction curve normalized to the $V$-band value steep.

The case shown in Figs.~\ref{fig:coag_shat_eta01} and \ref{fig:ext_coag_shat_eta01}
indicates that porosity evolution can have a dramatic impact on the shape of
extinction curves. Even if the grain size distribution becomes the one similar
to the initial grain size distribution at later stages of evolution,
the small porosity makes the extinction curves significantly steeper than
the initial one. In particular, the grains contributing to the optical extinction
have radii $a\sim \lambda /(2\upi )\sim 0.1~\micron$, where the porosity is the largest.
Therefore, if we consider the creation of porosity through the interplay between coagulation
and shattering,
the evolution of extinction curve could be qualitatively very different.
Moreover, the porosity depends on the grain radius; this dependence also creates
`higher-order' wavelength dependence of extinction curve.
As shown above, the wavelength range
where the extinction is reduced by porosity shifts to longer wavelengths if porosity is
developed in larger grains.

As shown in Section~\ref{subsec:coag_shat} (Fig.~\ref{fig:coag_shat_compact}),
if we consider compaction of shattered remnants, the filling factor increases a little
at $t\gtrsim 300$~Myr. We also calculated the evolution of extinction curve with
remnant compaction (now shown). Since coagulation is less efficient in this case,
the extinction curves are steeper than those shown in Fig.\  \ref{fig:ext_coag_shat_eta01}.
If we only see the extinction curve shape in the UV--optical,
a larger filling factor and a lower abundance of
large grains are degenerate.

\section{Conclusion}\label{sec:conclusion}

We formulate and calculate the evolution of grain size distribution and filling factor (porosity)
through coagulation and shattering in the ISM. We adopt the 2D Smoluchowski equation
to solve the distribution functions of
grain size and filling factor. To save the computational time, we only treat the
mean filling factor for each grain radius based on O09 and O12.
For coagulation, the transition from the hit-and-stick to
compaction regime are characterized and modelled by comparing the impact energy with
the rolling energy. Shattering is treated as
a formation mechanism of small compact fragments. For shattered remnants,
we basically assume the same porosity as the original grain but we also examine
the case where the volume equal to the colliding particle suffers perfect compaction.
We assume that coagulation and shattering are
hosted by the dense and diffuse ISM, respectively.

For the pure coagulation case (without shattering), the porosity develops around
$a\sim 0.01$--0.1 $\micron$, where a low filling factor of $\phi_m\sim 0.3$ is achieved.
However, when the porosity becomes significantly large, the major part of grains have already been
coagulated to $a>0.1~\micron$, where compaction occurs.
Therefore, the porosity little affects the evolution of grain size distribution if only coagulation
is present.
For the pure shattering case, we confirm that shattering tends to make the filling factors
asymptotically approach $\phi_m =1$ at $a\lesssim 0.01~\micron$, although those
at $a\gtrsim 0.03~\micron$ depend on the treatment of compaction for
shattered remnants.

Next, we examine the case where coagulation and shattering are both present.
This corresponds to a situation where we consider a wide enough area in the
ISM which contains both the dense and diffuse ISM, or where the time-scale of interest is
much longer than the exchange time of the ISM phases.
We find that the filling factor drops even below 0.1
around $a\sim 0.1~\micron$. The porosity evolution is sensitive to the
relative efficiency of coagulation to shattering, which is regulated by the dense gas
fraction, $\eta_\mathrm{dense}$. The filling factor tends to be small if shattering
is stronger (e.g.\ $\eta_\mathrm{dense}=0.1$). This is because shattering continues
to efficiently provide small grains, which are subsequently coagulated to form porous
grains. Thus, the interplay between shattering and coagulation is fundamentally
important as an origin of the porosity in the interstellar grains. For the case with
$\eta_\mathrm{dense}=0.1$, coagulation is activated in later stages (after porosity
develops) because the
high porosity enhances the grain cross-sections. Compaction is not efficient in this case
since the grain velocity is diminished by the increased porosity. Thus, large grains
are kept porous in this case.

We also calculate the evolution of extinction curve using the EMT. Porosity formed
as a result of coagulation enhances the UV and infrared extinction by $\sim$10--20 per cent.
As noted in previous studies, the extinction of porous silicate grains is suppressed
in the optical, and the wavelength range where the extinction is suppressed shifts towards
shorter wavelengths
if small grains are more abundant (or shattering is more efficient). The extinction is enhanced
at all wavelengths for amC in most of the cases.
A case with strong shattering ($\eta_\mathrm{dense}=0.1$) shows a recreation of
large grains at later stages as mentioned above. In this case, although the grain
size distribution itself is similar to the MRN distribution, the extinction curve shape
stays steep if we normalize the extinction to the $V$ band value. This
is because porosity makes the grains relatively `transparent' in the optical, while the
extinction is enhanced in the UV. Thus, the steepness of extinction curve is also
affected by the porosity evolution.

Although the predicted features in the extinction curves could be compared with
observations, our model developed in this paper is still premature
for detailed comparison. There are two necessary extensions of our modelling.
First, we could include other processes which also play an important role
in the evolution of grain size distribution; that is, dust production by stellar ejecta,
dust destruction by supernova shocks, and dust growth by the accretion of gas-phase metals.
Secondly, since the filling factor and the grain size distribution could be degenerate
in the resulting extinction curve shape, it is desirable to predict other independent
properties such as infrared emission SED and polarization. We emphasize that
the basic framework developed in this paper provides a basis on which we will extend
our predictions.

\section*{Acknowledgements}
 
We are grateful to S. Okuzumi, Y. Matsumoto, {and the anonymous referee} for useful comments.
HH thanks the Ministry of Science and Technology for support through grant
MOST 107-2923-M-001-003-MY3 and MOST 108-2112-M-001-007-MY3, and the Academia Sinica
for Investigator Award AS-IA-109-M02.
VBI acknowledges the support from the RFBR grant 18-52-52006 and
the SUAI grant FSRF-2020-0004.
{LP acknowledges support from the Programme National
`Physique et Chimie du Milieu Interstellaire' (PCMI) of CNRS/INSU with INC/INP
co-funded by CEA and CNES and from Action F\'{e}d\'{e}ratrice Astrochimie de
l'Observatoire de Paris. HH acknowledges the financial support and the hospitality
of l'Observatoire de Paris during his stay.}

\section*{Data Availability}

The data underlying this article are available in Figshare at
\url{https://doi.org/10.6084/m9.figshare.12917375.v1}.



\bibliographystyle{mnras}
\bibliography{/Users/hirashita/bibdata/hirashita}


\appendix

\section{Derivation of the basic equations}\label{app:derivation}

We derive equations (\ref{eq:rho}) and (\ref{eq:psi}) based on
O09. We start from the
2D Smoluchowski equation generalized to treat shattering as well as
coagulation:
\begin{align}
\lefteqn{\frac{\upartial f(\bmath{I}, t)}{\upartial t}
= -f(\bmath{I},\, t)\int K (\bmath{I};\,\bmath{I}')f(\bmath{I}',\, t)\,\mathrm{d}^2\bmath{I}'}
\nonumber\\
&+ \iint K(\bmath{I}_1;\, \bmath{I}_2)f(\bmath{I}_1,\, t)f(\bmath{I}_2,\, t)
\theta (\bmath{I};\, \bmath{I}_1,\, \bmath{I}_2)\,\mathrm{d}^2\bmath{I}_1\,
\mathrm{d}^2\bmath{I}_2,\label{eq:Smoluchowski}
\end{align}
where $f$ is the distribution function of $\bmath{I}\equiv (m,\, V)$ (grain mass
and volume, respectively) at time $t$, $K(\bmath{I}_1,\,\bmath{I}_2)$ is 
the collisional kernel (product of the collisional cross-section and the relative
velocity of the two colliding grains), $\theta (\bmath{I};\,\bmath{I}_1,\,\bmath{I}_2)$
is the distribution function of the produced grains by the collision, and the integration is
executed for all the relevant range of $\bmath{I}$, which is usually $[0,\,\infty ]\times [0,\,\infty ]$.
We distinguish the two colliding grains with subscipts 1 and 2, referred to as
grain 1 and 2, respectively [i.e. $\bmath{I}_1=(m_1,\, V_1)$ and $\bmath{I}_2=(m_2,\, V_2)$].
The normalization of $\theta$ is determined by
\begin{align}
\int m\theta (\bmath{I};\,\bmath{I}_1,\,\bmath{I}_2)\,\mathrm{d}^2\bmath{I}=m_1.
\end{align}
We note that when we consider the collision of grain 1 with grain 2,
we only consider the redistribution of $m_1$
(i.e.\ we separately treat the collision of grain 2 with grain 1).
This is why only $m_1$ enters
the normalization. Because of this, we do not have a factor of $1/2$
(which appears in O09's expression) before the
second term of the right-hand side in equation (\ref{eq:Smoluchowski}).
The two expressions are mathematically equivalent.

Now we take the zeroth moment of equation (\ref{eq:Smoluchowski}) for $V$, that is,
we integrate it for $V$. We adopt the following form for $\theta$ for simplicity and
for analytical convenience:
\begin{align}
\theta (\bmath{I};\,\bmath{I}_1,\,\bmath{I}_2)=
\bar{\theta}(m;\, m_1,\, m_2)\,\delta [V-V_{1+2}(m;\,\bmath{I}_1,\,\bmath{I}_2)],
\end{align}
where $\bar{\theta}$ describes the mass distribution function of
the grains formed after the collision between grains with $\bmath{I}_1$
and $\bmath{I}_2$, $\delta$ is Dirac's delta function, and $V_{1+2}$ describes the
volume of the grain formed from the collision between grains 1 and 2
(and the produced grain has a mass of $m$).
This expression assumes that $\bar{\theta}$ is independent of the volume
(or porosity) of the original grains and that the volume of the collisional product
is determined by $\bmath{I}_1$ and $\bmath{I}_2$ and is a function of $m$.
By integrating both sides of equation (\ref{eq:Smoluchowski}) for $V$, we obtain the
equation for the zeroth moment, $\tilde{n}(m,\, t)$, defined in equation (\ref{eq:n}).
We also take the first moment of equation (\ref{eq:Smoluchowski}); that is,
we multiply both sides of equation (\ref{eq:Smoluchowski}) by $V$ and integrate them for
$V$ to obtain the equation for $\bar{V}$ defined by equation (\ref{eq:V}).
The resulting moment equations are written as (see also O09)
\begin{align}
\lefteqn{\frac{\upartial \tilde{n}(m,\, t)}{\upartial t}=-\tilde{n}(m,\, t)\int\bar{K}(m;\, m_1)
\tilde{n}(m_1,\, t)\,\mathrm{d}m_1}\nonumber\\
& +\iint\bar{K}(m_1;\, m_2)\tilde{n}(m_1,\, t)\tilde{n}(m_2,\, t)\bar{\theta}(m;\, m_1,\, m_2)\,
\mathrm{d}m_1\,\mathrm{d}m_2,\label{eq:mom0}
\end{align}
\begin{align}
\lefteqn{\frac{\upartial \bar{V}(m,\, t)\tilde{n}(m,\, t)}{\upartial t}=-\tilde{n}(m,\, t)\int\overline{VK}(m;\, m_1)
\tilde{n}(m_1,\, t)\,\mathrm{d}m_1}\nonumber\\
& +\iint\overline{V_{1+2}K}(m_1;\, m_2)\tilde{n}(m_1,\, t)\tilde{n}(m_2,\, t)
\bar{\theta}(m;\, m_1,\, m_2)\,
\mathrm{d}m_1\,\mathrm{d}m_2,\label{eq:mom1}
\end{align}
where
\begin{align}
\bar{K}(m_1;\, m_2)\equiv\iint K(m_1,\, V_1;\, m_2,\, V_2)
\frac{f(m_1,\, V_1)}{\tilde{n}(m_1)}\frac{f(m_2,\, V_2)}{\tilde{n}(m_2)}\mathrm{d}V_1\mathrm{d}V_2,
\end{align}
\begin{align}
\overline{VK}(m;\, m_1)\equiv\iint VK(m,\, V;\, m_1,\, V_1)
\frac{f(m,\, V)}{\tilde{n}(m)}\frac{f(m_1,\, V_1)}{\tilde{n}(m_1)}\mathrm{d}V\mathrm{d}V_1,
\end{align}
\begin{align}
\overline{V_{1+2}K}(m; m_1,\, m_2) &\equiv \iint V_{1+2}(m;\, m_1,\, V_1,\, m_2,\, V_2)
K(m_1,\, V_1;\, m_2,\, V_2)\nonumber\\
&\times
\frac{f(m_1,\, V_1)}{\tilde{n}(m_1)}\frac{f(m_2,\, V_2)}{\tilde{n}(m_2)}\mathrm{d}V_1\mathrm{d}V_2.
\end{align}
The integration is performed for $[0,\,\infty ]\times [0,\,\infty]$.
The moment equations, in general, are not closed since a higher-order moment
always appears. To close this hierarchy, we adopt the same assumption as O09;
that is, the volume is replaced with the mean value at each $m$. Under this
assumption, the distribution function is written as
$f(m,\, V)=\tilde{n}(m)\,\delta [V-\bar{V}(m)]$. O09 refer to this approximation as
the volume-averaging approximation, and confirmed that it
gives a consistent result with the full solution of the 2D Smoluchowski equation
for their coagulation problem. Although there is no guarantee that this approximation is
valid for shattering, the increasing filling factor at small grain radii (Section~\ref{subsec:shat})
is at least qualitatively consistent with the evolution expected from
the production of compact fragments by shattering.

Adopting the above volume-averaging approximation, we obtain
\begin{align}
\bar{K}(m_1;\, m_2)=K[m_1,\,\bar{V}(m_1);\, m_2,\,\bar{V}(m_2)],
\end{align}
\begin{align}
\overline{VK}(m;\, m_1)=\bar{V}(m)\bar{K}(m;\, m_1),
\end{align}
\begin{align}
\overline{V_{1+2}K}(m;\, m_1,\, m_2)=V_{1+2}
[m;\, m_1,\,\bar{V}(m_1),\, m_2,\,\bar{V}(m_2)]\bar{K}(m_1;\, m_2).
\end{align}
We apply these relations to equations (\ref{eq:mom0}) and (\ref{eq:mom1}),
reorganize the notations as
$\bar{K}(m_1;\, m_2)=K_{m_1,m_2}$,
$V_{1+2}[m;\, m_1,\,\bar{V}(m_1),\, m_2,\,\bar{V}(m_2)]=(V_{1+2})_{m_1,m_2}^m$.
and multiply both sides in equation (\ref{eq:mom0}) by $m$.
Finally, introducing $\rho (m,\, t)\equiv m\tilde{n}(m,\, t)$ and
$\psi (m,\, t)\equiv \bar{V}(m,\, t)\tilde{n}(m,\, t)$, we
obtain equations (\ref{eq:rho}) and (\ref{eq:psi}).

\bsp	
\label{lastpage}
\end{document}